%% file: main.tex
\documentclass[conference]{IEEEtran}
\IEEEoverridecommandlockouts
\usepackage{cite}
\usepackage{amsmath,amssymb,amsfonts}
\usepackage{algorithmic}
\usepackage{graphicx}
\usepackage{textcomp}
\usepackage{xcolor}

\usepackage[ruled,lined]{algorithm2e}
\usepackage{amsthm}
\usepackage[normalem]{ulem}

\def\BibTeX{{\rm B\kern-.05em{\sc i\kern-.025em b}\kern-.08em
    T\kern-.1667em\lower.7ex\hbox{E}\kern-.125emX}}

\usepackage[colorinlistoftodos]{todonotes}
 \usepackage{cancel}

\input{macro}

\begin{document}
    \title{Minimizing Age of Processed Information in Wireless Networks}

 \author{\IEEEauthorblockN{Chanikarn~Nikunram*\IEEEauthorrefmark{2}, 
                          Wasin~Meesena*\IEEEauthorrefmark{3}, 
                          Stephen~John~Turner\IEEEauthorrefmark{2}, 
                          Sucha~Supittayapornpong\IEEEauthorrefmark{2}
            \thanks{
            \IEEEauthorrefmark{2}Vidyasirimedhi Institute of Science and Technology, Thailand.                
            \IEEEauthorrefmark{3}Carleton College, United States.
            *Both authors contributed equally to this research.}
            }}
\maketitle 
      \input{0-Abstract}
      

      \input{1-Introduction}

\input{3-SystemModel}

      \input{4-LowerBound}
      \input{5-Policies}
      \input{6-SimulationResult}

\input{7-Conclusion}

    \bibliographystyle{IEEEtran}
    \bibliography{reference}
    
   \input{8-Appendix}

\end{document}

%% file: macro.tex
\newcommand{\algonamepowertwo}{Power-2 }
\newcommand{\algonamebackoff}{Back-Off } 
\newcommand{\algonamemx}{Max-Weight }

\newcommand{\proctime}[1]{D_{{#1}}}

\newcommand{\procage}[2]{A^P_{{#1}}({#2})} 
\newcommand{\proctimePW}[1]{D^{PW}_{{#1}}}
\newcommand{\proctimeBO}[1]{D^{BO}_{{#1}}}

\newcommand{\notationprocessaoi}{AoIP} 
\newcommand{\textprocessaoi}{Age of  Information Processed} 

\newcommand{\AWSAoIabbre}{AWSAoIP} 
\newcommand{\AWSAoIBabbre}{AWSAoIU} 
\newcommand{\AWSAoI}[1]{J_P^{#1}} 
\newcommand{\AWSAoIB}[1]{J_U^{#1}} 
\newcommand{\textunprocessaoi}{Age of Information Unprocessed} 
\newcommand{\notationunprocessaoi}{AoIU} 
\newcommand{\agebs}[2]{A_{{#1}}^U({#2})} 
\newcommand{\textAWSprocessaoi}{Average Weighted sum $AoIP$} 

\newcommand{\pibackoff}{\pi^{BO} } 
\newcommand{\pimx}{\pi^{MW} }

\newcommand{\SrcSet}{\mathcal{N}}
\newcommand{\tx}[2]{x_{{#1}}({#2})}
\newcommand{\ifbusy}[2]{f_{{#1}}({#2})}
\newcommand{\policyclass}{\Pi}
\newcommand{\policyclassdf}{\Pi^{DF}}

\newcommand{\ratepolicy}[2]{r_{{#1}}^{#2}}
\newcommand{\lowerbound}{L_B}

\newcommand{\numpacket}[2]{V_{#1}[{#2}]} 
\newcommand{\interval}[2]{I_{#1}[{#2}]} 
\newcommand{\remainslot}[1]{R_{#1}} 
\newcommand{\rateop}[1]{r^{*}_{#1}}

\newcommand{\slotcandidate}[1]{\mathcal{P}_{{#1}}}
\newcommand{\basictime}[1]{B_{{#1}}}

\newcommand{\ifbusybackoff}[2]{b_{{#1}}({#2})}
\newcommand{\countdownname}{countdown time }
\newcommand{\countdown}[2]{C_{{#1}}({#2})}
\newcommand{\waitgapat}[1]{G_{{#1}}}
\newcommand{\firstzeroconnectedchainexample}{\Gamma}

\newcommand{\connectedchain}{Connected Chain }
\newcommand{\firstzeroconnectedchain}{{First Zero Connected Chain}}
\newcommand{\chainsetfzcg}{G^*}
\newcommand{\lengthchain}{L_C}

\newcommand{\whichnodesent}[1]{ S({#1})}

\newcommand{\networkstate}[1]{S({#1})}
\newcommand{\lfunction}[1]{L({#1})}

\newcommand{\figref}[1]{Figure~\ref{#1}}

\newcommand{\abs}[1]{\left|{#1}\right|}
\newcommand{\ceiling}[1]{\left\lceil{#1}\right\rceil}
\newcommand{\floor}[1]{\left\lfloor{#1}\right\rfloor}
\newcommand{\prtc}[1]{\left\{{#1}\right\}}
\newcommand{\prts}[1]{\left[{#1}\right]}
\newcommand{\prtr}[1]{\left({#1}\right)}

\newtheorem{corollary}{Corollary}
\newtheorem{lemma}{Lemma}
\newtheorem{theorem}{Theorem}

\DeclareMathOperator*{\argmax}{arg\,max}
\DeclareMathOperator*{\argmin}{arg\,min}
\DeclareMathOperator*{\mymod}{mod}

\newtheorem{innercustomthm}{Lemma}
\newenvironment{customthm}[1]
  {\renewcommand\theinnercustomthm{#1}\innercustomthm}
  {\endinnercustomthm}

%% file: 0-Abstract.tex

\begin{abstract}
The freshness of real-time status processing of time-sensitive information is crucial for several applications, including healthcare monitoring and autonomous vehicles. This freshness is considered in this paper for the system where unprocessed information is sent from sensors to a base station over a shared wireless network. The base station has a dedicated non-preemptive processor with a constant processing time to process information from each sensor. The age of processed information is the time elapsed since the generation of the packet that was most recently processed by a processor. Our objective is to minimize the average age of processed information over an infinite time-horizon. We first show that a drop-free policy simplifies the system without sacrificing optimality. From this simplification, we propose three transmission-scheduling policies with $2$-optimal guarantees for different requirements. A distributed Power-2 policy can be implemented without a central scheduler. With a central scheduler, both Back-Off and Max-Weight policies are near optimal with different advantages. The Back-Off policy guarantees a bound on the maximum age of processed information, while the Max-Weight policy achieves the lowest average age in simulation without the guarantee of bound. Simulation results confirm our theoretical findings.

\end{abstract}


%% file: 1-Introduction.tex

\section{Introduction}

\begin{figure}
    \centering
    \includegraphics[width=\linewidth]{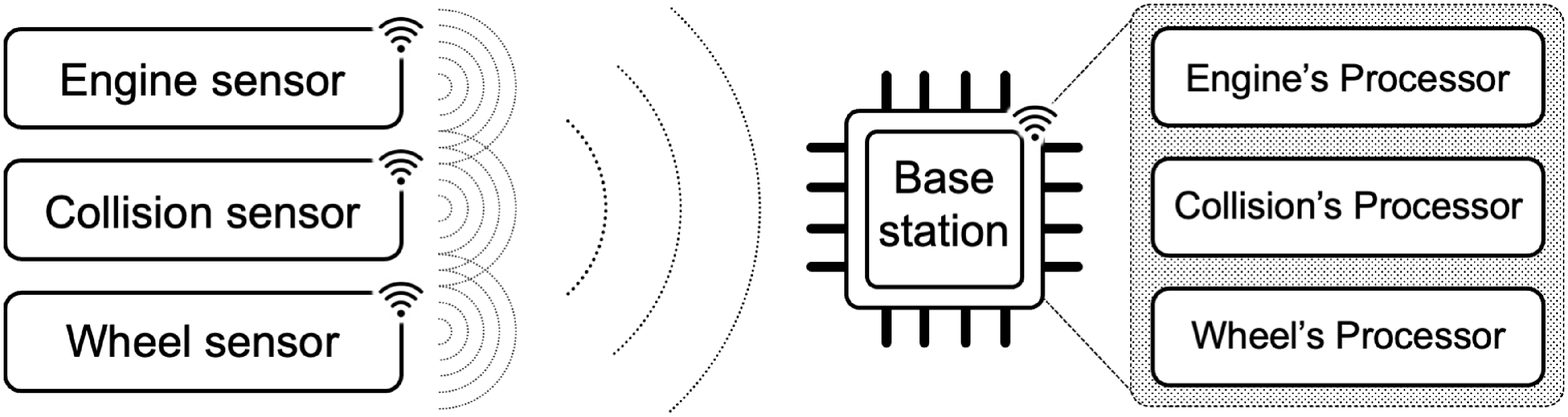}
    \caption{Time-sensitive information in a wireless network}
    \label{fig:simple_system}
\end{figure}

Real-time status processing finds various applications, including healthcare monitoring \cite{healthcare} 
and autonomous vehicles \cite{autocarjointICC,eautocaricc}. These applications collect real-time information from sensors over wireless communication, process the obtained information at a base station, and utilize the processed information for making decisions as illustrated in \figref{fig:simple_system}. For example, an autonomous vehicle needs to process information in real-time to drive safely. The vehicle's control system receives data wirelessly from sensors, and data from each sensor is then processed by a dedicated processor.

To ensure that systems always make informed decisions based on up-to-date data, the freshness of processed information becomes one of the key factors.
The freshness of information can be rigorously modeled by the \emph{age of information} (AoI) (\cite{survey, book} and references therein), which is the time elapsed since the generation of the packet that the base station most recently received. AoI has been extensively studied from queueing analysis, including single-server queues \cite{howoften}, multi-source queues \cite{mutisource}, non-preemptive queues \cite{nonpreemptive}, and parallel-server queues \cite{parallelserver}, to scheduling problems, such as throughput constraints \cite{throughtputmodiano}, power constraints \cite{power, energy2022}, single-hop \cite{broadcastrandom,broadcastmodiano,wirelessmodiano, mat}, multi-hop \cite{multihopPri,funda,multihop}, both offline and online policies \cite{design}. Moreover, \cite{unified, gen,joint2021} have investigated AoI from the perspective of joint sampling and scheduling policies. However, those studies focus on the \emph{age of information unprocessed} (AoIU) of packets that have just arrived at the base station. In some circumstances, the newly received information must be processed before the systems can use it. Thus, we must incorporate this processing time into scheduling procedures.

In this paper, the \emph{age of information processed} (AoIP), which is the time elapsed since the generation of the packet that was most recently processed by a processor, is modeled to quantify the freshness of information for the real-time status processing. As shown in \figref{fig:simple_system}, time-sensitive and unprocessed information is transmitted from sensors to a base station over a shared wireless channel. To process information from each sensor, the base station possesses a dedicated non-preemptive processor with a constant processing time, and each processor may process one packet at a time. Each processor is said to have fresher information when a new packet has been fully processed--rather than when a new packet arrives at the base station. Our goal is to develop scheduling policies for a resource-constrained wireless channel that minimizes the average age of processed information over an infinite time-horizon.
While the \notationprocessaoi \ can better capture the aspect of processed information, it is challenging to derive, compared to the \notationunprocessaoi. Thus, we concentrate on a subclass of scheduling policies, namely drop-free policies, that make the analysis of the average age of processed information more tractable. We show that this subclass of policies does not sacrifice any optimality and derives a lower performance bound to the optimal average sum of \notationprocessaoi. Afterward, we present three drop-free scheduling policies, called the \algonamepowertwo policy, the \algonamebackoff policy, and the \algonamemx policy. They all achieve $2$-optimal performance guarantees with different advantages.

To elaborate, the \algonamepowertwo policy is low-complexity and fully distributed. 
On the other hand, with central scheduler employing a greedy approach, the \algonamebackoff policy can significantly outperform the \algonamepowertwo policy in terms of the average age. Furthermore, the \algonamebackoff policy guarantees a bound on the maximum age of processed information. Lastly, the \algonamemx policy is a centralized policy that uses a Lyapunov Optimization technique. In simulation results, it provides the average age closest to the optimal. However, the \algonamemx policy cannot guarantee a bound on the maximum age of processed information.

The contributions of this paper are threefold:
\begin{itemize}
     \item We model the age of processed information in a wireless network with processors and show that a class of drop-free policies tractably minimizes \notationprocessaoi~ without sacrificing any optimality.
     
     
     
     \item We analyze a lower performance bound of the system and design three drop-free scheduling policies with 2-optimal guarantees. 
     
     \item We simulate these policies and compare them with the lower performance bounds to confirm our analytical results and show that the \algonamemx and \algonamepowertwo policies perform near-optimal.
     
\end{itemize}
This paper is organized as follows. The system model is formally presented in Section \ref{sec:model}.  
Section \ref{sec:dropfree} introduces the drop-free policies and analyzes their lower performance bounds. Three policies with different advantages are proposed and analyzed in Section \ref{sec:policy}. In Section \ref{sec:simulation}, these policies are simulated to evaluate their performances, followed by the conclusion in Section \ref{sec:conclusion}. 

%% file: 3-SystemModel.tex
\section{System Model}
\label{sec:model}

\subsection{Network and Processors}

\begin{figure}
    \centering
    \includegraphics[width=\linewidth]{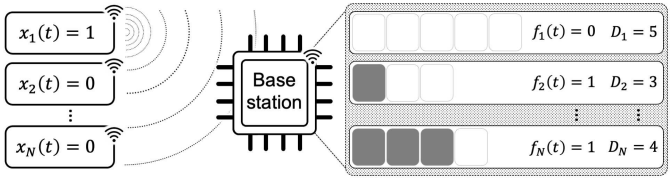}
    \caption{A system model consists of $N$ sensors and dedicated processors with different processing times. Each gray block represents a progress bar. For example, the total processing time of processor 2 is $3$ slots, and the current progress is $1/3$.}
    \label{fig:system_model}
\end{figure}

Consider a wireless network system with $N$ nodes attempting to send packets, containing \emph{time-sensitive, unprocessed information}, to their corresponding $N$ processors at a base station, as shown in \figref{fig:system_model}. Each node and its corresponding processor are indexed by the index set $\SrcSet = \prtc{1, \dotsc, N}$. For example, sensors in an autonomous vehicle send the measured data wirelessly to a central control system, the base station in our model, with a dedicated processor for each sensor.

Time is slotted, $t \in \prtc{1, 2, \dotsc}$. In each slot $t$, at most one node can transmit a packet to the base station over a wireless channel due to interference, which is similar to \cite{funda,unified,mat, gen, throughtputmodiano,broadcastrandom,design,Energy2021,wirelessmodiano}. 
Let $\tx{i}{t} $ be a decision variable that equals $1$ when node $i$ transmits a packet to the base station in slot $t$, and $\tx{i}{t}=0$ otherwise. Hence, the \emph{interference constraint} is defined by
\begin{align}\label{interferenceconstraint}
\sum_{i \in \SrcSet} \tx{i}{t} \leq 1, \quad \forall t \in \prtc{1, 2, \dotsc}.
\end{align}
We assume that every transmission is always successful when the interference constraint is satisfied \cite{mat, gen,broadcastrandom}.

In this paper, we consider the class of all scheduling policies, denoted by $\Pi$, that make transmission decision $\prtc{\tx{i}{t}}_{i \in \SrcSet}$ every slot such that the interference constraint is satisfied. Henceforth, we assume all transmissions satisfy the interference constraint. Every transmitted packet is assumed to take one slot to arrive at the base station \cite{funda,broadcastrandom,throughtputmodiano, mat, power}. When a packet arrives at the base station, it is redirected immediately to a corresponding processor.

All processors are non-preemptive. Each processor $i$ can process one packet at a time, and the processing time takes $\proctime{i}$ consecutive slots for any constant $\proctime{i} \in \mathbb{N}$. Therefore, when a new packet for processor $i$ arrives at time $t$ and the processor is idle, not processing any packet, the packet will be completely processed at the end of slot $t + \proctime{i} - 1$ and will be considered as \emph{processed} information in slot $t + \proctime{i}$. However, if processor $i$ is busy (not idle), the newly arrived packet will be \emph{dropped} and is no longer considered for processing. We call this situation \emph{wasteful transmission}. Note that the system could introduce queues to store newly-arrived packets while processors are not idle. However, later we will show in Section~\ref{sec:dropfree} that those queues are unnecessary for optimality.

\subsection{Wasteful Transmission}

To prevent wasteful transmission, where a transmitted packet is dropped, we introduce $\ifbusy{i}{t}$ as an indicator variable that equals $1$ if a transmission from node $i$ in slot $t$ will be wasteful, and $\ifbusy{i}{t}=0$ otherwise. We assume all processors are initially idle, so $\ifbusy{i}{1}=0, \forall i\in \SrcSet$. 
When node $i$ transmits a packet in slot $t$ knowing that the transmission will not be wasteful, i.e., $\tx{i}{t}=1$ and $\ifbusy{i}{t}=0$, at the beginning of slot $t+1$ the processor $i$ will be \textit{busy} processing this transmitted packet until the end of slot $t+\proctime{i}$ as illustrated in \figref{fig:system_model}. This implies that $\ifbusy{i}{\tau}=1$ for $\tau \in \prtc {t+1, \dotsc ,t+\proctime{i}-1}$ and $\ifbusy{i}{t+\proctime{i}} = 0$. In particular, any packet from node $i$ transmitted during $\prtc{t+1, \dotsc,t+\proctime{i}-1}$ will be dropped, but a packet transmitted at time $t+\proctime{i}$ will get processed right after the previous processing completes.

If node $i$ does not transmit in slot $t$ and $\ifbusy{i}{t}=0,$ then $\ifbusy{i}{t+1}=0$. 
However, if node $i$ transmits a packet in slot $t$ knowing that the transmission will be wasteful, i.e. $\tx{i}{t}=1$ and $\ifbusy{i}{t}=1,$ then the transmitted packet will be dropped in slot $t+1$ right after it has been redirected to processor $i.$ Also, this wasteful transmission will not affect the values of $\ifbusy{i}{t}.$ The dynamic of $\ifbusy{i}{t}$ is illustrated in Figure \ref{fig:example}.


\begin{figure}
    \centering
 \includegraphics[width=\linewidth]{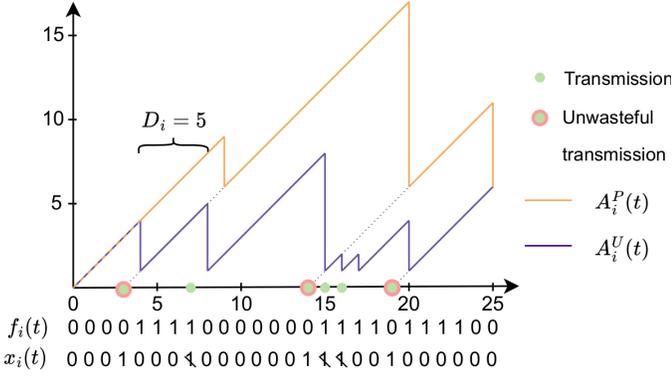} 
    \caption{  The green dots represent transmissions from node $i$. The pink dots represent unwasteful transmissions when those packets that would not be dropped at the base station are generated and transmitted to the base station. The crossed $\tx{i}{t}$ represents a wasteful transmission. A wasteful transmission affects $\agebs{i}{t}$ but does not affect $\procage{i}{t},$ as show in slots $7, 15$, and $16$.}
    \label{fig:example}%
\end{figure}

\subsection{Age of Information}

The Age of Information (AoI) represents how fresh the information received by the base station is. In this paper, we extend AoI to two metrics, considering unprocessed and processed information. They are formally defined with their relationship below. We also assume that every packet is generated right before transmission, and its age starts at zero.


The \textunprocessaoi~ (\notationunprocessaoi) of node $i$ at time $t$ represents how old the transmitted packet that the base station most recently received from node i is.
This \notationunprocessaoi ~is denoted by a positive integer $\agebs{i}{t}$. When node $i$ transmits a packet at time $t$, i.e., $\tx{i}{t}=1$, and other nodes do not, this newly-generated packet from node $i$ will arrive at the base station in slot $t+1$, and $\agebs{i}{t+1}=1$ because the packet is generated right before the transmission and takes one slot to arrive. When there is no transmission from node $i$, the AoIU gets older by one slot, and $\agebs{i}{t+1}=\agebs{i}{t}+1$. The $\agebs{i}{t+1}$ is illustrated in Figure \ref{fig:example} and is written as follows
\begin{align}\label{evole-of-aoi-at-bs}
    \agebs{i}{t+1}=\begin{cases} 
   1        & \text{, if $\tx{i}{t}= 1$;}\\
      \agebs{i}{t}+1    & \text{, otherwise.} \\
   \end{cases}
\end{align}

To measure the \notationunprocessaoi~ of the entire system when a policy $\pi$ is employed, we consider the \emph{Average Weighted Sum of AoIU} (\AWSAoIBabbre) as follows, for every $\pi \in \Pi$,
\begin{align}\label{awsaoib-def}
    \AWSAoIB{\pi}=\lim_{T\to \infty} \frac{1}{NT} \sum_{t=1}^{T}\sum_{i \in \SrcSet} w_i\agebs{i}{t}
\end{align}
where $w_i$ is a positive weight of node $i$.

The \textprocessaoi~ (\notationprocessaoi) of node $i$ at time $t$ represents how old the packet that processor $i$ most recently finished processing is. This AoIP is denoted by a positive integer $\procage{i}{t}$. When processor $i$ has just finished processing a packet at the end of slot $t$, we know that $ \procage{i}{t+1} = \proctime{i} + \agebs{i}{t - \proctime{i} +1} = \proctime{i} + 1$ because i) the processing time is $\proctime{i}$ and ii) the received packet at the base station is not dropped and is immediately processed. In particular, this situation happens only when the packet is transmitted at time $t - \proctime{i}$, and the transmission is unwasteful, i.e., $\tx{i}{t-\proctime{i}}= 1$ and $\ifbusy{i}{t-\proctime{i}} = 0$. When there is no processing completion of processor $i$ at time $t$, the \notationprocessaoi~ gets older by one slot, and $\procage{i}{t+1}=\procage{i}{t}+1$. The $\procage{i}{t+1}$ is illustrated in Figure \ref{fig:example} and is written as follows 
\begin{align}\label{aoi}
    \procage{i}{t+1}=\begin{cases} 
    \proctime{i}+1 & \text{, if $\tx{i}{t-\proctime{i}}= 1$ and} \\ 
    \phantom{ }  & \text{   \ \ \ \ $\ifbusy{i}{t-\proctime{i}} = 0$;}\\
    \procage{i}{t}+1 & \text{, otherwise.} \\
   \end{cases}
\end{align}
To measure \notationprocessaoi~ of the entire system when a policy $\pi$ is employed, we consider the \emph{Average Weighted Sum of AoIP} (\AWSAoIabbre) as follows, for every $\pi \in \Pi$, 
\begin{align}\label{awsaoi-def}
    \AWSAoI{\pi}=\lim_{T\to \infty} \frac{1}{NT} \sum_{t=1}^{T}\sum_{i \in \SrcSet} w_i\procage{i}{t}
\end{align}
with the same weight $w_i$ defined in \eqref{awsaoib-def}.

The initial \notationunprocessaoi~ and \notationprocessaoi~ can be any arbitrary positive integers. For simplicity, we assume that $\agebs{i}{1} = \procage{i}{1} = 1$ for all $i \in \SrcSet$.

\subsection{System Objective}

In this paper, we develop scheduling policies that minimize the \AWSAoIabbre~ in \eqref{awsaoi-def} under the interference constraint in \eqref{interferenceconstraint} as follows  
\begin{subequations}
\label{optimizprob}
\begin{align}
  \AWSAoI{*}=&\min_{\pi \in \Pi} \bigg\{ \lim_{T\to \infty} \frac{1}{NT} \sum_{t=1}^{T}\sum_{i \in \SrcSet} w_i\procage{i}{t} \bigg\} \label{optimizprob:cost}\\
        & \text{s.t. } \ \sum_{i \in \SrcSet} \tx{i}{t} \leq 1, \quad \forall t \in \prtc{1, 2, \dotsc}, \label{optimizprob:const1}
\end{align}
\end{subequations}
where $\AWSAoI{*}$ is the minimum \AWSAoIabbre~ achieved under a scheduling policy that satisfies the interference constraint. While we aim to minimize the \AWSAoIabbre, the evolution of $\procage{i}{t}$ in \eqref{aoi} is too cumbersome to deal with, compared to $\agebs{i}{t}$ in \eqref{evole-of-aoi-at-bs}. The next section considers a special class of policies that render our analysis tractable.

Note that, while the scheduling problem in \eqref{optimizprob} is deterministic, computing the optimal policy offline needs to search over a policy space, which is exponentially large with respect to $N$ and $\prod_{i\in \SrcSet} \proctime{i}$. For example, let $L_{i}{(t)}$ be the state of processor $i$ at time $t$. At time $t,$ a policy $\pi$ could map the current system state, $\prtc{\procage{i}{t}, L_{i}{(t)}}_{i\in \SrcSet}$, to a scheduling decision at time $t+1$. In this case, searching for an optimal policy $\pi$ for large $N$ and $\prod_{i\in \SrcSet} \proctime{i}$ is intractable. Furthermore, implementing the policy requires a large memory space to store the policy mapping.



%% file: 4-LowerBound.tex

\section{Drop-free Policies and Preliminary}
\label{sec:dropfree}


We observe that wasteful transmission consumes network resources without improving \AWSAoIabbre. Therefore, we consider a \emph{drop-free scheduling policy} that avoids scheduling a transmission from nodes that will cause a packet drop, once the packet arrives at the base station. We denote the class of drop-free policies by $\policyclassdf$. Note that $\policyclassdf\subseteq \policyclass$. Next, we introduce some useful properties of drop-free policies, derive a lower performance bound, and prove a preliminary lemma.



\subsection{Drop-free Policies}
\label{ssec:drop_free_policy}

For each node $i$, we begin with decomposing the first $T$ slots into intervals between transmissions. Let $\numpacket{i}{T}$ be the total number of packets transmitted by node $i$ within these $T$ slots, i.e., $\sum_{t=1}^{T} \tx{i}{t}=\numpacket{i}{T}$. Let $\interval{i}{m}$ be the number of slots between the $(m-1)$th and $m$th packet deliveries from node $i$ for all $m \in \prtc{1,2, \dotsc,\numpacket{i}{T}}$, and let $\remainslot{i}$ be the number of remaining slots after the last transmission to slot $T$. Therefore, we have
\begin{align} \label{timeslot-interval}
    T=\sum_{m=1}^{\numpacket{i}{T}} \interval{i}{m} +\remainslot{i}, \quad \forall i \in \SrcSet.
\end{align}
Next, we show that drop-free policies lead to a simple relationship between unprocessed and processed ages.

\begin{lemma}\label{relationbetweenages}
For any drop-free scheduling policy, the relationship between $\agebs{i}{t}$   in \eqref{evole-of-aoi-at-bs} and  $\procage{i}{t}$ in (\ref{aoi}) always follows:
\begin{align}\label{relation}
    \procage{i}{t+\proctime{i}}=\agebs{i}{t}+\proctime{i}, \quad  \forall t \geq \interval{i}{1}+1, ~ \forall i \in \SrcSet.
\end{align}
\end{lemma}
\begin{proof}
We will prove this by induction. Let $\pi$ be a drop-free policy. Since $\tx{i}{t}=1$ implies that $\ifbusy{i}{t}=0, $ the evolution of   $\procage{i}{t}$ in (\ref{aoi}) can be rewritten as 
\begin{align}\label{aoi2}
     \procage{i}{t+1}=\begin{cases} 
      \proctime{i}+1        & \text{, if $\tx{i}{t-\proctime{i}}= 1$;}\\
      \procage{i}{t}+1    & \text{, otherwise.} \\
   \end{cases}
\end{align}

\textbf{Basic Step:} Since $\tx{i}{\interval{i}{1}}=1,$ we know from \eqref{evole-of-aoi-at-bs} and \eqref{aoi2} that 
\begin{align*}
\agebs{i}{\interval{i}{1}+1}+\proctime{i}= 1+\proctime{i}=\procage{i}{\interval{i}{1}+\proctime{i}+1}.
\end{align*}
 
\textbf{Inductive Step:} 
Suppose $k \geq \interval{i}{1}+1$ is an integer such that 
\begin{equation}\label{inductionstep}
	\procage{i}{k+\proctime{i}} = \agebs{i}{k}+\proctime{i}.
\end{equation} 
For $k+1$ and $\tx{i}{k}= 1$, we have from \eqref{evole-of-aoi-at-bs} and \eqref{aoi2} that
\begin{align*} 
	\agebs{i}{k+1}+\proctime{i}=\proctime{i}+1 = \procage{i}{k+\proctime{i}+1}.
\end{align*}
For $k+1$ and $\tx{i}{k}= 0$, we have that
\begin{align*}
	\agebs{i}{k+1}+\proctime{i} & = \agebs{i}{k}+1+\proctime{i} \notag\\
	& = \procage{i}{k+\proctime{i}} + 1 = \procage{i}{k+\proctime{i}+1}, 
\end{align*}
where the first equality uses \eqref{evole-of-aoi-at-bs}, the second equality uses the induction hypothesis \eqref{inductionstep}, and the last equality uses \eqref{aoi2}. Hence, the equality in \eqref{relation} holds by the proof of induction.
\end{proof}

The implication of Lemma \ref{relationbetweenages} is that the relationship between \notationunprocessaoi~ and \notationprocessaoi~ becomes much simpler under drop-free policies. The next lemma considers the relationship between \AWSAoIBabbre~ and \AWSAoIabbre.

\begin{lemma}[\AWSAoIBabbre \ and \AWSAoIabbre]
\label{awsaoi-and-awsaoib}
For any system with $\prtc{w_i, \proctime{i}}_{i\in \SrcSet}$, the following holds for every drop-free scheduling policy, $\pi \in \policyclassdf$,  
\begin{align}\label{awsaoi-and-awsaoib-equation}
   \AWSAoI{\pi} = \AWSAoIB{\pi} +\frac{1}{N}\sum_{i \in \SrcSet} w_i\proctime{i},   
\end{align}
where $\AWSAoI{\pi}$ and $\AWSAoIB{\pi}$ are defined repectively in \eqref{awsaoi-def} and \eqref{awsaoib-def}.
\end{lemma}
\begin{proof}
Consider a drop-free policy $\pi \in \policyclassdf$. Manipulating the \AWSAoIabbre~ in $\eqref{awsaoi-def}$ gives 
\begin{align*}
      \AWSAoI{\pi}&=\lim_{T\to \infty} \frac{1}{NT}\sum_{i \in \SrcSet} w_i \Bigg[ \sum_{t=1}^{\interval{i}{1}+\proctime{i}} \procage{i}{t}\\
      &\quad \quad\quad\qquad\quad \quad + \sum_{t=\interval{i}{1}+\proctime{i}+1}^{T} \procage{i}{t} \Bigg].
\end{align*}
Knowing the first term above disappears after applying the limit $T\to\infty$, we apply Lemma \ref{relationbetweenages} as $\pi$ is drop-free and obtain 
\begin{align*}
      \AWSAoI{\pi} &=\lim_{T\to \infty} \frac{1}{NT}\sum_{i \in \SrcSet} w_i\sum_{t=\interval{i}{1}+1}^{T-\proctime{i}} \prtr{ \agebs{i}{t}+\proctime{i} } \\
      & = \lim_{T\to \infty} \frac{1}{NT}\sum_{i \in \SrcSet} w_i \Bigg[\sum_{t=1}^{T-\proctime{i}} \prtr{ \agebs{i}{t}+\proctime{i} }\\ 
      & \quad \quad \quad \quad \quad \quad \quad\qquad- \sum_{t=1}^{\interval{i}{1}} \prtr{ \agebs{i}{t}+\proctime{i} } \Bigg]
\end{align*}
Knowing the second term in the RHS disappears in the limit, we use the definition of \AWSAoIBabbre~ in \eqref{awsaoib-def} to prove \eqref{awsaoi-and-awsaoib-equation} from
\begin{align*}
      \AWSAoI{\pi} & = \lim_{T\to \infty} \frac{1}{NT}\sum_{i \in \SrcSet} w_i \sum_{t=1}^{T-\proctime{i}} \agebs{i}{t}   + \frac{1}{N}\sum_{i \in \SrcSet} w_i\proctime{i}.
\end{align*}
\end{proof}

Next, we argue that considering the class of drop-free policies does not affect the optimal  \AWSAoIabbre~ in \eqref{optimizprob:cost}. 
\begin{lemma}[Drop-free Optimal Scheduler] \label{optimal-drop-free} For any system with $\prtc{w_i, \proctime{i}}_{i\in \SrcSet}$, there exists a drop-free optimal scheduler. 
\end{lemma}
\begin{proof}
Assume that $S$ is an optimal scheduler that knows which node to transmit in each slot to achieve the optimal \AWSAoIabbre. If the scheduler $S$ causes wasteful transmissions, then we can simply construct a new scheduler $S^*$ by replicating $S$ but without transmitting those packets that would have been dropped according to the scheduler $S$. By doing so, $S^*$ is a drop-free scheduler, and its \AWSAoIabbre \ is not higher than the optimal \AWSAoIabbre \ under the optimal schedule $S$. Thus, $S^*$ is a drop-free optimal scheduler.
\end{proof}

Note that even if a processor is preemptive, meaning it has the capability to drop a packet it is processing, there still exists an optimal drop-free policy because we can tweak an optimal policy for preemptive processors by not scheduling those packets that would have been preempted later. In particular, the tweaked policy is drop-free and achieves the same optimality for the system with preemptive processors.

Combining Lemma \ref{awsaoi-and-awsaoib} and Lemma \ref{optimal-drop-free}, we have that 
\begin{equation}\label{lower-bound2}
    \AWSAoI{*} = \min_{\pi \in \policyclassdf } \AWSAoI{\pi} = \min_{\pi \in \policyclassdf }\AWSAoIB{\pi} +\frac{1}{N}\sum_{i \in \SrcSet} w_i\proctime{i}. 
\end{equation}
This simplifies our optimization only to minimize \AWSAoIBabbre~ over drop-free policies. However, finding the optimal \AWSAoIBabbre~ is challenging, so we mitigate this by deriving a lower performance bound and later show that our policies in Section \ref{sec:policy} achieve the bound.



\subsection{Lower Performance Bound}


We consider the \emph{long-term transmission rate} and derive its property before proceeding to the lower bound. The long-term transmission rate of node $i$ under policy $\pi$ is defined by, for every $\pi \in \Pi$,
\begin{align}\label{long-term-transmissio-rate}
    \ratepolicy{i}{\pi}=\lim_{T\to \infty} \frac{1}{T} \sum_{t=1}^{T} \tx{i}{t}=\lim_{T\to \infty}  \frac{\numpacket{i}{T}}{T}.
\end{align}

\begin{lemma}[Maximum Transmission Rates]
\label{upper-bound-rate}
For any system with $\prtc{w_i, \proctime{i}}_{i\in \SrcSet}$ and for any drop-free policy $\pi \in \policyclassdf,$ the long-term transmission rate from each node is at most the reciprocal of the corresponding processing time. In particular, the following holds
\begin{align}\label{rate-bound}
\ratepolicy{i}{\pi} \leq \frac{1}{\proctime{i}}, \quad \forall i \in \SrcSet.
\end{align}
\end{lemma}
\begin{proof}
Fix node $i \in \SrcSet$. Consider the first  $T$ slots. Since $\pi$ is a drop-free policy, the inter-delivery times during $T$ slots must be at least $\proctime{i},$ i.e., 
\begin{align*}
    \interval{i}{m} \geq \proctime{i}, \quad \forall i \in \SrcSet \text{ and } m \in\prtc{ 2,3, \dotsc, \numpacket{i}{T}}.
\end{align*}
Applying the above inequality to \eqref{timeslot-interval}, we get
\begin{align*}
    T&=\sum_{m=1}^{\numpacket{i}{T}} \interval{i}{m} +\remainslot{i} \geq \prtr{\numpacket{i}{T}-1}\proctime{i}.
\end{align*}
Manipulating the inequality above, applying the limit $T \to \infty$, and using the definition of the long-term transmission rate in \eqref{long-term-transmissio-rate} yields
\begin{align*}
\frac{1}{\proctime{i}} \geq  \lim_{T\to \infty} \prts{\frac{\numpacket{i}{T}}{T} -\frac{1}{T} }= \ratepolicy{i}{\pi}.
\end{align*}
\end{proof}

\begin{theorem}[Lower bound]
\label{lower-bound-thm}
For any system with $\prtc{w_i, \proctime{i}}_{i\in \SrcSet}$, the optimal \AWSAoIabbre~ $\AWSAoI{*}$ has the lower bound $\lowerbound$ that is the solution of the following optimization problem, i.e., $\AWSAoI{*} \geq \lowerbound$ and
\begin{subequations}
\label{bound-problem}
\begin{align}
     \lowerbound =  & \min_{\pi \in \policyclassdf} \prtc{ \frac{1}{N}\sum_{i \in \SrcSet} w_i \prtr{\proctime{i}+ \frac{1}{2}+\frac{1}{2\ratepolicy{i}{\pi}}} } \label{bound-problem:obj}\\
                & \quad \text{s.t.}   \sum_{i \in \SrcSet} \ratepolicy{i}{\pi} \leq 1 \label{bound-problem:const1}\\
                & \qquad 0<\ratepolicy{i}{\pi}  \leq \frac{1}{\proctime{i}}, \quad \forall i \in \SrcSet. \label{bound-problem:const2}
\end{align}
\end{subequations}
\end{theorem}
\begin{proof}
Consider a drop-free policy $\pi \in \policyclassdf$ that satisfies the interference constraint throughout the first $T$ slots. Fix node $i$, and recall that the notations $\numpacket{i}{T}, \interval{i}{m}$, and $\remainslot{i}$ are introduced in Section \ref{ssec:drop_free_policy} for equality \eqref{timeslot-interval}. For each interval $\interval{i}{m}$, the unprocessed age $\agebs{i}{t}$ evolves as $\prtc{1,2, \dotsc , \interval{i}{m}}$ according to the dynamic of $\agebs{i}{t}$ in \eqref{evole-of-aoi-at-bs}. Similarly, the age evolves as $\prtc{1,2, \dotsc, \remainslot{i}}$ during the remaining $\remainslot{i}$ slots. Therefore, the time-average age of information unprocessed of node $i$ during the first $T$ slots is as follows
%
%
\begin{align}\label{derive-aoi}
    \frac{1}{T} \sum_{t=1}^{T} \agebs{i}{t}
    &=\frac{1}{T} \prts{ \sum_{m=1}^{\numpacket{i}{T}}\frac{(\interval{i}{m}+1)\interval{i}{m}}{2}+\frac{(\remainslot{i}+1)\remainslot{i}}{2} } \\
    &=\frac{1}{2T} \prts{ \sum_{m=1}^{\numpacket{i}{T}} \interval{i}{m}^2+ \remainslot{i}^2 } +\frac{1}{2}. \nonumber
\end{align}
Applying Cauchy's Inequality to the above equality gives
\begin{align*}
    \frac{1}{T} \sum_{t=1}^{T} \agebs{i}{t} &\geq \frac{1}{2T}\frac{ \prts{ \sum_{m=1}^{\numpacket{i}{T}} \interval{i}{m}+ \remainslot{i} }^2}{\numpacket{i}{T}+1} +\frac{1}{2} \\
    & =\frac{T}{2(\numpacket{i}{T}+1)} +\frac{1}{2}.
\end{align*}
Applying the limit $T\to \infty$ to the above and using the definition of the long-term transmission rate \eqref{long-term-transmissio-rate}, we obtain the lower bound of the time-average age of information unprocessed for node $i$ as 
\begin{align*}
     \lim_{T \to \infty}\frac{1}{T} \sum_{t=1}^{T} \agebs{i}{t} 
     & \geq  \frac{1}{ 2 \lim_{T \to \infty} \prts{ \frac{\numpacket{i}{T}}{T}+\frac{1}{T} } } +\frac{1}{2} \\
     &\geq \frac{1}{2 \ratepolicy{i}{\pi}} +\frac{1}{2}.
\end{align*}
Using the bound above, we can bound the \AWSAoIBabbre~ in \eqref{awsaoib-def} by  
\begin{align}\label{weightedsumbound}
    \AWSAoIB{\pi} &=   \frac{1}{N}\sum_{i \in \SrcSet} w_i \lim_{T\to \infty}\frac{1}{T}\sum_{t=1}^{T}\agebs{i}{t} \nonumber\\
    &\geq \frac{1}{N} \prts{ \sum_{i \in \SrcSet} \frac{ w_i}{2 \ratepolicy{i}{\pi}} + \sum_{i \in \SrcSet}\frac{ w_i}{2} }.
\end{align}
The lower bound in \eqref{weightedsumbound} is a function of long-term transmission rates. Using the definition of these rates in \eqref{long-term-transmissio-rate} and summing over $i\in \SrcSet$, we have 
\begin{align}\label{sum-rate-constrint2}
   \sum_{i\in \SrcSet} \ratepolicy{i}{\pi}&=  \lim_{T\to \infty} \frac{1}{T} \sum_{i\in \SrcSet}\sum_{t=1}^{T} \tx{i}{t}=  \lim_{T\to \infty} \frac{1}{T}\sum_{t=1}^{T}  \sum_{i\in \SrcSet}\tx{i}{t} \leq 1
\end{align}
where the last inequality holds because the policy satisfies the interference constraint in \eqref{interferenceconstraint}.
%
%
Minimizing both sides of \eqref{weightedsumbound} over drop-free policies and using \eqref{sum-rate-constrint2} and \eqref{rate-bound} as the constraints give
\begin{align}\label{bs-vs-bound}
   \min_{\pi \in \policyclassdf} \AWSAoIB{\pi} \geq  \min_{\pi \in \policyclassdf}  \frac{1}{N} \sum_{i \in \SrcSet} w_i \prtr{ \frac{1}{2 \ratepolicy{i}{\pi}} +\frac{1}{2} }
\end{align}
 where $\sum_{i \in \SrcSet} \ratepolicy{i}{\pi} \leq 1$ and  $  0<\ratepolicy{i}{\pi}  \leq \frac{1}{\proctime{i}}$  for all $i \in \SrcSet.$ Finally, using \eqref{bs-vs-bound} and \eqref{lower-bound2} proves the theorem and \eqref{bound-problem}.
\end{proof}

Intuitively, Theorem \ref{lower-bound-thm} provides a lower performance bound of an optimal drop-free policy. It is later used to analyze the performances of three scheduling algorithms in Section \ref{sec:policy}. Note that our lower bound in \eqref{bound-problem} differs from \cite{throughtputmodiano,gen} in that i) it involves processing constraints and ii) it is under a class of drop-free policies.

Furthermore, it is possible to show that this lower bound also holds for another system with queues placed in front of the processors, where processors can be either preemptive or non-preemptive. In other words, if some drop-free scheduling policy achieves the bound, that policy is also near-optimal for the system with queues. 
This claim can be proven by considering a rate of processed packets.
Since this rate cannot exceed the transmission rate and packets get older when being queued, we can derive a similar bound as in \eqref{bound-problem}.


The bound in \eqref{bound-problem} can be made explicit by finding the set of \emph{optimal rates} $\prtc{\rateop{i}}_{i \in \SrcSet}$ that minimizes \eqref{bound-problem:obj} under constraints \eqref{bound-problem:const1} and \eqref{bound-problem:const2}. These optimal rates can be found from solving the convex optimization in \eqref{bound-problem} by any off-the-shelf solver, for example, \textsf{CVX} \cite{cvx}. 
Substituting the optimal rates $\prtc{\rateop{i}}_{i \in \SrcSet}$ into \eqref{bound-problem:obj} gives
\begin{align}\label{bound}
    \AWSAoI{*} \geq \lowerbound=  \frac{1}{N} \sum_{i \in \SrcSet} w_i \prtr{ \frac{1}{2 \rateop{i}} +\frac{1}{2} +\proctime{i} }.
\end{align}
Furthermore, when $\sum_{i \in \SrcSet} \frac{1}{\proctime{i}} \leq 1,$ it is easy to see that $\rateop{i} = \frac{1}{\proctime{i}}$ for all node $i$, and we have
\begin{align}\label{bound-special}
 \AWSAoI{*} \geq\lowerbound =\frac{1}{N} \sum_{i \in \SrcSet} w_i \prtr{ \frac{3}{2}\proctime{i}+ \frac{1}{2} }.
\end{align}

\subsection{Preliminary of Performance Guarantee}
To guarantee the performance of a scheduling policy $\pi \in \policyclass$, 
we say $\pi$ is $k$-optimal if $\AWSAoI{\pi} \leq k \times \AWSAoI{*}$ where $ \AWSAoI{*} = \min_{\pi' \in \policyclass} \AWSAoI{\pi'}$. We first prove the following lemma, which will be used in the next section.

 
 \begin{lemma}[inter-deliver-optimal]\label{inter-deliver-optimal}
When a scheduling policy has the property that $\ceiling{\frac{1}{\rateop{i}}} \leq \interval{i}{m}\leq 2\ceiling{\frac{1}{\rateop{i}}}$ for every $i \in \SrcSet$ and $m \in \mathbb{N},$ the policy is $2$-optimal. Moreover, it is $4/3$-optimal if $\sum_{i\in \SrcSet} \frac{1}{\proctime{i}} \leq 1.$
 \end{lemma}

 \begin{proof}
Considering the first $T$ slots with the similar derivation in \eqref{derive-aoi} and using the condition that $\remainslot{i}$, which is lower than $\interval{i}{\numpacket{i}{T}+1}$, and $\interval{i}{m} $ do not exceed $ 2\ceiling{\frac{1}{\rateop{i}}},$ we have 
\begin{align*}
    \frac{1}{T} \sum_{t=1}^{T} \agebs{i}{t} 
    &=\frac{1}{2T}\prts{\sum_{m=1}^{\numpacket{i}{T}} \interval{i}{m}^2+ \remainslot{i}^2 }+\frac{1}{2} \nonumber\\
    & \leq  \frac{ 2\ceiling{\frac{1}{\rateop{i}}} }{2T}\prts{\sum_{m=1}^{\numpacket{i}{T}} \interval{i}{m}+ \remainslot{i} }+\frac{1}{2}\\
    & =\ceiling{\frac{1}{\rateop{i}}}+\frac{1}{2}.
\end{align*}
Since $\interval{i}{m} \geq \ceiling{\frac{1}{\rateop{i}}} \geq \frac{1}{\rateop{i}} \geq \proctime{i},$ $\pi$ is a drop-free policy. Therefore, by Lemma \ref{awsaoi-and-awsaoib} and the inequality above, we obtain
 \begin{align}
 \label{ineq-aoi}
\AWSAoI{\pi} &= \AWSAoIB{\pi}+\sum_{i\in \SrcSet}w_i\proctime{i} \nonumber\\
     &=  \lim_{T \to \infty}\frac{1}{T} \sum_{t=1}^{T} \sum_{i \in \SrcSet} w_i\agebs{i}{t}+\sum_{i\in \SrcSet}w_i\proctime{i} \nonumber \\
     &\leq \sum_{i\in \SrcSet} w_i\prtr{\ceiling{\frac{1}{\rateop{i}}}+\frac{1}{2} +\proctime{i}}.
\end{align}
Since  $\ceiling{\frac{1}{\rateop{i}}} \leq \frac{1}{\rateop{i}}+\proctime{i}$ as $\proctime{i} \geq 1$, the inequality \eqref{ineq-aoi} yields that $\pi$ is 2-optimal according to \eqref{bound} as
 \begin{align*}
\AWSAoI{\pi} 
   & \leq \sum_{i\in \SrcSet} w_i\prtr{\frac{1}{\rateop{i}}+\frac{1}{2}+2\proctime{i}} \\
& \leq 2\sum_{i \in \SrcSet} w_i \prtr{\proctime{i}+ \frac{1}{2}+\frac{1}{2\rateop{i}}}= 2\AWSAoI{*}.
\end{align*}
%
Also, if $\sum_{i \in \SrcSet} \frac{1}{\proctime{i}} \leq 1$, then  $\rateop{i}=\frac{1}{\proctime{i}}$ for all $i\in \SrcSet$ and, the inequality \eqref{ineq-aoi} yields that $\pi$ is 4/3-optimal according to \eqref{bound-special} and
\begin{align*}
\AWSAoI{\pi}  
& \leq \sum_{i\in \SrcSet} w_i\prtr{\proctime{i}+\frac{1}{2}+\proctime{i}} \\
& \leq   \frac{4}{3}  \sum_{i \in \SrcSet} w_i \prtr{\frac{3}{2}\proctime{i}+ \frac{1}{2}}\leq \frac{4}{3} \AWSAoI{*}.
\end{align*}
\end{proof}
 

%% file: 5-Policies.tex

\section{Scheduling Policies}
\label{sec:policy}

In this section, we present three scheduling policies, namely, \algonamepowertwo, \algonamebackoff, and Max-Weight, with their performance guarantees.

\subsection{\algonamepowertwo Policy}
\label{sec:power2}

The \algonamepowertwo policy minimizes \AWSAoIabbre~ with a $2$-optimal guarantee. It is distributed, cyclic, and has low complexity. The policy is inspired by the work in \cite{mat} for a different problem that guarantees the upper bound of AoI without minimizing \AWSAoIBabbre \ explicitly. Here, we present our policy and derive its performance guarantee.

Without loss of generality, node indices are rearranged such that $ \rateop{1}\geq\rateop{2}\geq \dotsc \geq \rateop{N}$. We define $\proctimePW{i}=2^{\ceiling{\log_2\prtr{1/\rateop{i}}}}$ as a fixed inter-delivery time for source node $i$. Note that $ \proctimePW{1}\leq\proctimePW{2}\leq \dotsc \leq \proctimePW{N}$. Then, for each node $i$, we determine a \emph{basic time} $\basictime{i}$, in which every packet from node $i$ is transmitted in slot $t$ where $t \equiv \basictime{i} \mymod \proctimePW{i}$. Intuitively, $\basictime{i}$ is the first slot node $i$ transmits a packet and it keeps transmitting every $\proctimePW{i}$ slots. Therefore, node $i$ only needs know $\basictime{i}$ and $\proctimePW{i}$ to make transmission decisions independently from other nodes, assuming the starting time is synchronized.



The procedure for determining the basic time $\prtc{\basictime{i}}_{i \in \SrcSet}$ that the policy is drop-free and satisfies the interference constraint in \eqref{interferenceconstraint} is summarized in Algorithm \ref{alg:Power2} with an example provided below.
\begin{algorithm}

\caption{Procedure for determining basic time $\prtc{\basictime{i}}$}\label{alg:Power2}
\SetKwInOut{Input}{Input}\SetKwInOut{Output}{Output}
\Input{$\prtc{\proctimePW{i}}_{i\in \SrcSet}$}
\Output{$\prtc{\basictime{i}}_{i\in \SrcSet}$}

 \For{$i \in \prtc{1, \dotsc, N}$}{
    $\slotcandidate{i} = \prtc{ t \in \prtc{1,2, \dotsc ,\proctimePW{i} } | t \not\equiv \basictime{j} \mymod{\proctimePW{j}}, \forall j < i }$\\
    $\basictime{i} = \min \slotcandidate{i} $\\
  }
\end{algorithm}



\textbf{Example: } A system consisting of three nodes with $\prtc{\proctime{i}}_{i=1}^3=\prtc{2,2,4}$ and $\prtc{w_i}_{i=1}^3=\prtc{20,5,1}$. Solving problem \eqref{bound-problem} obtains $\prtc{\rateop{i}}_{i=1}^3 \approx \prtc{\frac{1}{2}, \frac{1}{2.9} ,\frac{1}{6.5}}$, and we set $\prtc{\proctimePW{i}}_{i=1}^3=\prtc{2^{\ceiling{\log_2\prtr{1/\rateop{i}}}}}_{i=1}^3 = \prtc{2, 4 ,8}$. Starting from the first node, we have $\slotcandidate{1}=\prtc{1,2}$ and $\basictime{1}=1$. Next, we have $\slotcandidate{2}=\prtc{2,4}$ and $\basictime{2}=2$. Lastly, we have $\slotcandidate{3}=\prtc{4,8}$ and $\basictime{3}=4$. Therefore, node $1$ transmits a packet in slots $\prtc{1,3,5,7, \dotsc}$ without communicating with the other nodes. Similarly, node $2$ transmits in slots $\prtc{2,6,10,14, \dotsc}$, and node $3$ transmits in slots $\prtc{4,12,20, \dotsc}$.


Now, we show that Algorithm \ref{alg:Power2} always constructs the basic times $\prtc{\basictime{i}}_{i \in \SrcSet}$ before showing that the interference constraint in \eqref{interferenceconstraint} is satified by the transmissions generated by these basic times.
\begin{theorem}[Existence of Basic Times]
For any system with $\prtc{w_i, \proctime{i}}_{i \in \SrcSet}$, Algorithm \ref{alg:Power2} always leads to non-empty $\slotcandidate{i}$ for all $i \in \SrcSet$.
\end{theorem}
\begin{proof}
Let $\prtc{\proctimePW{i}}_{i\in N}$ be the result of a given network setting. We use induction to show that $\slotcandidate{i}$ is non-empty for all $i \in \SrcSet$.


\textbf{Basic Step:} Since $\proctimePW{i}\geq 1,$ we have that $\slotcandidate{1}=\prtc{1,2, \dotsc,\proctimePW{1}}$ and is non-empty.

\textbf{Inductive Step:} Let $k \in \prtc{1,2, \dotsc,N-1}$ such that all $\slotcandidate{k}$ are non-empty. We will show that $\slotcandidate{k+1}$ is non-empty as follows.
\begin{align} \label{inequality-p_k}
    &\abs{\slotcandidate{k+1}} \nonumber\\
    & \quad = \Bigg| \prtc{1, 2, \dotsc, \proctimePW{k+1} } \Big/ \bigcup_{j=1}^k \prtc{ t \in \prtc{1, \dotsc, \proctimePW{k+1} } }\nonumber\\
    & \quad\quad\quad\quad\quad\quad\quad\quad\quad\quad\quad\quad\quad\quad \big| t \equiv \basictime{j} \mymod \proctimePW{j} \Bigg| \\
    & \quad \geq \proctimePW{k+1} - \sum_{j=1}^{k}\Bigg|\prtc{ t \in \prtc{1,2,\dotsc,\proctimePW{k+1} }}\nonumber\\ 
    & \quad\quad\quad\quad\quad\quad\quad\quad\quad\quad\quad\quad\quad\quad \big| t \equiv \basictime{j} \mymod \proctimePW{j} \Bigg|.
\end{align}
Because all $\proctimePW{i}$ are power of $2$ and $\proctimePW{1} \leq \dotsc \leq \proctimePW{k+1}$, we know that $\proctimePW{k+1}$ is divisible by $\proctimePW{j}$ for every $j \leq k.$ It follows that   
\begin{align*}
\abs{\prtc{ t \in \prtc{1,2, \dotsc ,\proctimePW{k+1} } \big| t\equiv \basictime{j} \mymod{\proctimePW{j}}} }=\frac{\proctimePW{k+1} }{\proctimePW{j}}.
\end{align*}
Applying the property above to \eqref{inequality-p_k} gives
 \begin{align*}
       \abs{\slotcandidate{k+1}} \geq \proctimePW{k+1} -\sum_{j=1}^{k} \frac{\proctimePW{k+1} }{\proctimePW{j}}
       =\proctimePW{k+1}\prtr{1-\sum_{j=1}^{k} \frac{1}{\proctimePW{j}}}.
 \end{align*}
 Using the facts that $\frac{1}{\proctimePW{i}} \leq 2^{-\log_2\prtr{1/\rateop{i}}} = \rateop{i}$ 
 and $\sum_{i \in \SrcSet} \ratepolicy{i}{*} \leq 1$, as $\prtc{\rateop{i}}_{i \in \SrcSet}$ satisfies \eqref{bound-problem:const1}, we have 
  \begin{align*}
       \abs{\slotcandidate{k+1}}
       \geq \proctimePW{k+1}\prtr{1-\sum_{j=1}^{k} \rateop{j}}  
      & \geq \proctimePW{k+1}\prtr{ \rateop{k+1}} \\
       & \geq \proctimePW{k+1} \frac{1}{\proctimePW{k+1}}=1.
 \end{align*}
Therefore, $\slotcandidate{k+1}$ is non-empty, which proves the theorem.
\end{proof}

\begin{theorem}[No Concurrent Transmissions]
For any system with $\prtc{w_i, \proctime{i}}_{i \in \SrcSet}$, the \algonamepowertwo policy satisfies the interference constraint in \eqref{interferenceconstraint}.

\end{theorem}

\begin{proof}
Let $\prtc{\proctimePW{i}}_{i\in N}$ be the result of a given network setting. Suppose that the \algonamepowertwo policy violates the interference constraint. There exist $i, j \in \SrcSet$ in which $i > j$ and $t \in \mathbb{N}$ such that $t \equiv \basictime{i} \mymod \proctimePW{i}$ and $t \equiv \basictime{j} \mymod \proctimePW{j}$, as nodes $i$ and $j$ are scheduled to transmit in the same slot. However, since $\proctimePW{i}$ is divisible by $\proctimePW{j},$ the fact that $t\equiv \basictime{i} \mymod{\proctimePW{i}}$ leads to $t\equiv \basictime{i} \mymod{\proctimePW{j}}.$ It follows that $\basictime{j} \equiv \basictime{i} \mymod \proctimePW{j}$. However, by the procedure for constructing $\basictime{i}$ in Algorithm \ref{alg:Power2}, the fact that $\basictime{i}\not\equiv \basictime{j} \mymod{\proctimePW{j}}$ leads to a contradiction. Therefore, the policy satisfies the interference constraint.
\end{proof}



We finally state the performance guarantee of the \algonamepowertwo policy.

\begin{theorem}[Performance Guarantee of \algonamepowertwo]
For any system with $\prtc{w_i, \proctime{i}}_{i \in \SrcSet}$, the \algonamepowertwo policy is $2$-optimal, and it is $4/3$-optimal if $\sum_{i\in \SrcSet} \frac{1}{\proctime{i} }\leq 1$.
\end{theorem}

\begin{proof}

To determine the \AWSAoIabbre~ of the \algonamepowertwo policy over an infinite time horizon, we can simply ignore the first inter-delivery time of every node. According to the policy, $\interval{i}{m}= 2^{\ceiling{\log_2\prtr{1/\rateop{i}}}}$ for every $i \in \SrcSet$ and $m \in \prtc{2, 3, 4, \dotsc}$. 
It holds that
\begin{align*}
    &\interval{i}{m}=2^{\ceiling{\log_2\prtr{1/\rateop{i}}}}\leq 2\cdot2^{\log_2\prtr{1/\rateop{i}}}\leq 2\ceiling{1/\rateop{i}}\\
 \text{and} \quad 
    &\interval{i}{m}=2^{\ceiling{\log_2\prtr{1/\rateop{i}}}}\geq \ceiling{2^{\log_2\prtr{1/\rateop{i}}}}= \ceiling{1/\rateop{i}}.
\end{align*}
Applying Lemma \ref{inter-deliver-optimal} proves the theorem.
\end{proof}

Intuitively, the \algonamepowertwo policy utilizes a set of fixed inter-deliver times to achieve the $2$-optimal guarantee with simple and distributed implementation. However, this comes at a cost of under-utilization of network resources, i.e., no transmission even when some processor is idle. The next policy improves on this issue.


\subsection{\algonamebackoff Policy}

The \algonamebackoff policy employs a greedy approach to select a node for transmission without causing wasteful transmissions. This is done by restricting the inter-delivery time to be at least $\proctimeBO{i} = \ceiling{1/\rateop{i}}$ for every node $i \in \SrcSet$. Later, we show that the policy is $2$-optimal and outperforms the \algonamepowertwo policy in simulations. In particular, it has bounded maximum $\notationprocessaoi$.


The \algonamebackoff~ policy maintains a list of candidate nodes for drop-free transmission in every slot. Let $\ifbusybackoff{i}{t}$ be an indicator variable that equals $0$ if node $i$ is a candidate for slot $t$. If $\ifbusybackoff{i}{t}$ equals $1$, node $i$ is \textit{backed off} from the list. We initialize $\ifbusybackoff{i}{t}=0 \ \forall i\in \SrcSet$ and $\forall t \in\mathbb{N}$, and then $\ifbusybackoff{i}{t}$ is updated dynamically as follows. If node $i$ transmits during slot $t$, it will back off from the list for $\proctimeBO{i}-1$ slots. That is $\ifbusybackoff{i}{\tau} = 1$ where $\tau \in \prtc{t+1, t+2, \dotsc, t+\proctimeBO{i}-1}$. Since $ \ceiling{1/\rateop{i}}\geq \proctime{i},$ the \algonamebackoff~ policy is drop-free.


There could be multiple candidate nodes in a slot. Therefore, we introduce a \textit{\countdownname} $\countdown{i}{t}$ as a variable that quantifies how urgent node $i$ should transmit in slot $t$. This $\countdown{i}{t}$ is initialized by $\countdown{i}{1} = \proctimeBO{i}$ and evolves as follows
\begin{align}\countdown{i}{t}=\begin{cases} 
      \proctimeBO{i} & , t> \proctimeBO{i} \text{ and }\tx{i}{t-\proctimeBO{i}}=1; \\
      \infty &, \ifbusybackoff{i}{t} = 1; \\
      \countdown{i}{t-1}-1 &, \text{otherwise.}
   \end{cases}\label{countdown-def}
\end{align}
That is, when node $i$ is backed-off, the countdown time is infinite. If it has just stopped being backed-off, then $\countdown{i}{t}$ is set to $\proctimeBO{i}.$ If the node is not backed-off and it does not transmit a packet, then the \countdownname reduces by $1.$ Intuitively, the lower $\countdown{i}{t}$ is, the more urgent node $i$ needs to transmit a packet in slot $t$. 
%
%
Thus, in each slot $t$, the \algonamebackoff policy schedules a transmission from node $j$ where
\begin{align*}
   j = \argmin_{i \in \SrcSet : \ifbusybackoff{i}{t}=0} \countdown{i}{t}.
\end{align*}
Also, if all nodes are backed off, $\ifbusybackoff{i}{t}=1$ for every $i\in \SrcSet$, nothing is transmitted during slot $t$. The \algonamebackoff policy is summarized in Algorithm \ref{alg:backoff}. 

\textbf{Example:} Table \ref{tab:example_case} illustrates the \algonamebackoff policy using the same setup in Section \ref{sec:power2}. Given $\prtc{\proctime{i}}_{i=1}^3=\prtc{2,2,4}$ and $\prtc{w_i}_{i=1}^3=\prtc{20,5,1}$, we have $\prtc{\rateop{i}}_{i=1}^3 \approx \prtc{\frac{1}{2}, \frac{1}{2.9} ,\frac{1}{6.5}},$ and $\prtc{\proctimeBO{i}}_{i=1}^3$ $=$ $\prtc{\ceiling{1/\rateop{i}}}_{i=1}^3 $ $= \prtc{2, 3 ,7}.$ We initialize the countdown times with $\prtc{\countdown{i}{1}}_{i=1}^3=\prtc{2, 3 ,7}$. Although processor $3$ takes $\proctime{3} = 4$ slots to process information and a transmission in the $8$th slot by node $3$ is not wasteful, $\ifbusy{3}{8}=0$, node $3$ is still backed-off until $t=10$. The countdown time is reset later in the $11$th slot, $\countdown{3}{11}=7$.

\begin{algorithm}
\caption{\algonamebackoff Policy}\label{alg:backoff}
\SetKwInOut{Input}{Input}\SetKwInOut{Output}{Output}
\Input{$\prtc{w_i, \proctime{i}, \proctimeBO{i}, \agebs{i}{1} }_{i\in \SrcSet}$}
\Output{$ \prtc{\prtc{\tx{i}{t}}_{i\in \SrcSet} }_{t\in\prtc{1,2, \dotsc}}$}
Initialize $\ifbusybackoff{i}{t} \gets 0, \countdown{i}{1} \gets \proctimeBO{i}, \forall i\in \SrcSet, \forall t\in\prtc{1,2, \dotsc}$\\
%

 \For{$t \in \prtc{1, 2, \dotsc}$}{
 	Initialize $\tx{i}{t} \gets 0, \forall i\in \SrcSet$\\
	\If{some nodes are not being backed-off at time $t$}{
   		Let $ j= \argmin_{i \in \SrcSet : \ifbusybackoff{i}{t}=0} \countdown{i}{t}$\\
   
   		Set $\tx{j}{t} \gets 1$\\
   
   
    		Set $\ifbusybackoff{j}{\tau} \gets 1, \forall \tau \in\prtc{t+1,t+2,\dotsc, t+\proctimeBO{i}-1}$\\
  }
  Update $\agebs{i}{t+1}, \procage{i}{t+1}$, and $\countdown{i}{t+1}$ according to \eqref{evole-of-aoi-at-bs}, \eqref{aoi}, and \eqref{countdown-def}\\
  }
\end{algorithm}

\begin{table}
\resizebox{\columnwidth}{!}{%
\begin{tabular}{|c|c|c|c|c|c|c|c|c|c|c|c|c|}
\hline
t       & 1 & 2 & 3 & 4 & 5 & 6 & 7 & 8 & 9 & 10 & 11 & 12 \\ \hline
$C_1(t)$ &$\cancel{2}$&$\infty$&$\cancel{2}$&$\infty$&$\cancel{2}$&$\infty$&  $\cancel{2}$ & $\infty$  &   $\cancel{2}$ &  $\infty$  &  $\cancel{2}$  &   $\infty$     \\ 
$C_2(t)$ &  3 &$\cancel{2}$&$\infty$&$\infty$&3&  $\cancel{2}$ & $\infty$  & $\infty$  &  3 &  $\cancel{2}$  & $\infty$   &  $\infty$    \\ 
$C_3(t)$ &  7 & 6 & 5  &$\cancel{4}$&$\infty$&$\infty$&$\infty$&$\infty$&$\infty$& $\infty$   &  7  &  $\cancel{6}$     \\ \hline
$\ifbusybackoff{3}{t}$  &  0 & 0  &  0  &  0 &  1 &  1 &  1 & 1 &  1 &  1  &  0  &  0     \\ \hline
$\ifbusy{3}{t}$  &  0 & 0  &  0  &  0 &  1 &  1 &  1 &  0 &  0  &  0  &  0  &  0     \\ \hline
\end{tabular}
}
\\
\caption{An example of the \algonamebackoff policy}
 \label{tab:example_case}
\end{table}


The performance of the \algonamebackoff policy can be analyzed by showing that every countdown time never reaches a zero as stated in the following lemma. Note that we set $\proctimeBO{i} = \ceiling{1/\rateop{i}}$ for all $i \in \SrcSet$, so $\sum_{i=1}^N \frac{1}{\proctimeBO{i}}\leq 1$ always holds.

\begin{lemma}[No-zero]
\label{No-zero}
If $\sum_{i=1}^N \frac{1}{\proctimeBO{i}}\leq 1$, then we have that $\countdown{i}{t}\geq 1$ for every $i\in \SrcSet$ and every $t \in \prtc{1,2, \dotsc}$. 
\end{lemma}

\begin{proof}
Since this proof requires defining new definitions and proving other three lemmas, it is provided in the Appendix. 
The high-level idea is as follows. We use the proof of contradiction by first assuming that there exists a non-stop transmission period such that some countdown reaches $0$.
However, due to the countdown times and back-off periods, the number of transmissions during this non-stop transmission period from each node is limited. As a result, the maximum total number of transmissions during this period is less than the length of the period, suggesting that there must be a slot with no transmission during the non-stop period and contradicting the assumption that the period is non-stop.
\end{proof}

We then use Lemma \ref{No-zero} to prove the following guarantee.

\begin{theorem}[Performance Guarantee of \algonamebackoff]
For any system with $\prtc{w_i, \proctime{i}}_{i \in \SrcSet}$, the \algonamebackoff policy is $2$-optimal, and it is $4/3$-optimal if $\sum_{i\in \SrcSet} \frac{1}{\proctime{i} }\leq 1$.
\end{theorem}

\begin{proof}
Since the back-off period is $\proctimeBO{i}-1=\ceiling{1/\rateop{i}}-1$, the interdelivery time $\interval{i}{m}$ is at least $\ceiling{1/\rateop{i}}$. From Lemma \ref{No-zero}, the countdown times are always positive, so the inter-delivery time $\interval{i}{m}$ is no longer than the sum of back-off period and the longest countdown period, i.e., $2\ceiling{1/\rateop{i}} - 1$. Therefore, we have $\ceiling{1/\rateop{i}} \leq \interval{i}{m}\leq 2\ceiling{1/\rateop{i}}$, and the policy is drop-free. Invoking Lemma \ref{inter-deliver-optimal} proves the theorem.
\end{proof}



\begin{corollary}[Bounded Maximum AoIP of \algonamebackoff]
\label{max-age}
The \algonamebackoff~ policy has bounded maximum AoIP satisfying
$$\procage{i}{t}\leq 2\ceiling{\frac{1}{\rateop{i}}}+\proctime{i}-1 \quad \forall i\in \SrcSet, \forall t \geq \interval{i}{1}+\proctime{i}+1.$$
\end{corollary}

\begin{proof}

Since the \algonamebackoff policy is drop-free, we can write $\procage{i}{t}$ as $\agebs{i}{t-\proctime{i}}+\proctime{i}$ for every $t \geq \interval{i}{1}+\proctime{i}+1$ by Lemma \ref{relationbetweenages}. Since, the maximum inter-delivery time is at most $2\ceiling{1/\rateop{i}}-1$ and every packet takes one slot to arrive at the base station, we have $\agebs{i}{t-\proctime{i}} \leq 2\ceiling{1/\rateop{i}}-1$ and
\begin{align*}
\procage{i}{t}=\agebs{i}{t-\proctime{i}}+\proctime{i}\leq 2\ceiling{\frac{1}{\rateop{i}}} +\proctime{i}-1.
\end{align*}
\end{proof}

The \algonamebackoff policy still has unused slots while some processors are idle. The next section considers a greedy scheduling policy that always makes unwasteful transmission when possible.

\subsection{Max-Weight Policy}

The Max-Weight policy is a drop-free policy that requires a central scheduler. It is derived from a Lyapunov Optimization technique \cite{book, throughtputmodiano}. We first derive the policy and then prove its performance.

The network state is defined by $\networkstate{t}=\prtc{\agebs{i}{t}, \ifbusy{i}{t}}_{i\in \SrcSet}$, where $\agebs{i}{t}$ and $\ifbusy{i}{t}$ are \notationunprocessaoi~ and the wasteful transmission indicator defined in Section \ref{sec:model}. We consider the linear Lyapunov function
\begin{align}\label{lyapunov}
    \lfunction{\networkstate{t}} =\frac{1}{N} \sum_{i\in \SrcSet} \tilde \alpha_i \agebs{i}{t}
\end{align}
where every $\tilde{\alpha_i}$ is some positive constant.
The Lyapunov drift is defined by
\begin{align} \label{lyapunov-drift}
     \Delta(\networkstate{t})= \lfunction{\networkstate{t+1}}- \lfunction{\networkstate{t}}.
\end{align} 
The evolution of $\agebs{i}{t+1}$ in \eqref{evole-of-aoi-at-bs} can be rewritten as
\begin{align}\label{linear-aoi}
    \agebs{i}{t+1}= \tx{i}{t} +\prtr{1-\tx{i}{t}}\prtr{\agebs{i}{t}+1}.
\end{align} 
Substituting \eqref{linear-aoi} and \eqref{lyapunov} to \eqref{lyapunov-drift}, we get 
\begin{align}\label{drift-derive}
    \Delta(\networkstate{t}) &= \frac{1}{N}\sum_{i\in \SrcSet} \tilde \alpha_i\prts{ \agebs{i}{t+1}- \agebs{i}{t} } \nonumber\\
    &=-\frac{1}{N}\sum_{i\in \SrcSet} \tx{i}{t} \tilde \alpha_i\agebs{i}{t}+\frac{1}{N}\sum_{i\in \SrcSet} \tilde \alpha_i.
\end{align} 
 
The Max-Weight policy minimizes the Lyapunov drift in \eqref{drift-derive} to reduce the progress of AoIU in every slot. Since the only controllable variable is the transmission $\tx{i}{t}$, the policy selects node $i$ that i) causes no wasteful transmission, $\ifbusy{i}{t}=0$, and ii) has the largest coefficient, $ \tilde \alpha_i\agebs{i}{t}$. In particular, node $j$ is selected for transmission when 
$$
j= \argmax_{i \in \SrcSet: \ifbusy{i}{t}=0} \tilde \alpha_i \agebs{i}{t}.
$$
The Max-Weight policy with $\tilde \alpha_i =w_i/\rateop{i}$ for every $i\in \SrcSet$ is summarized in Algorithm \ref{alg:MaxWeight}. Recall that $\ratepolicy{i}{*}$ is the optimal rate of the problem in \eqref{bound-problem}. The performance of Max-Weight is proven in Theorem \ref{thm:MaxWeight}.

\begin{algorithm}
\caption{Max-Weight Policy}\label{alg:MaxWeight}
\SetKwInOut{Input}{Input}\SetKwInOut{Output}{Output}
\Input{$\prtc{w_i, \rateop{i}, \agebs{i}{1} }_{i\in \SrcSet}$}
\Output{$ \prtc{\tx{i}{t}}_{i\in \SrcSet, t\in\prtc{1,2, \dotsc}}$}
Initialize $\ifbusy{i}{t} \gets 0, \forall i\in \SrcSet, \forall t \in \prtc{1, 2, \dotsc}$\\


\For{$t \in \prtc{1, 2, \dotsc}$}{
	Initialize $\tx{i}{t} \gets 0, \forall i\in \SrcSet$\\
	 \If{no wasteful transmission is possible at time $t$}{
  		Let $j = \argmax_{i \in \SrcSet : \ifbusy{i}{t}=0} \frac{w_i}{\rateop{i}} \agebs{i}{t}$\\
		Set $\tx{j}{t} \gets 1$\\
		Set $\ifbusy{j}{\tau} \gets 1, \forall \tau \in\prtc{t+1,t+2, \dotsc, t+\proctime{j}-1}$\\
	}
  	Update $\agebs{i}{t+1}$ and $\procage{i}{t+1}$ according to \eqref{evole-of-aoi-at-bs} and \eqref{aoi} 
}

\end{algorithm}

%
\begin{theorem}[Performance Guarantee of Max-Weight]
\label{thm:MaxWeight}
For any system with $\prtc{w_i, \proctime{i}}_{i \in \SrcSet}$ and corresponding optimal rate $\prtc{\rateop{i}}_{i \in \SrcSet}$, the Max-Weight policy with $\tilde \alpha_i=w_i/\rateop{i}$ for all $i \in\SrcSet$ is $2$-optimal, and it is $5/3$-optimal if $\sum_{i\in \SrcSet} \frac{1}{\proctime{i}} \leq 1.$
\end{theorem}
\begin{proof}
Let $\tx{i}{t}$ represent the decision variable with respect to the Max-Weight policy, $\pimx$. Since the sum of the optimal rate is at most $1$ as in \eqref{sum-rate-constrint2} and the policy selects the node with the highest $\tilde \alpha_i\agebs{i}{t}$, we have 
\begin{align}\label{max-weight-bound}
    \sum_{i \in \SrcSet : \ifbusy{i}{t}=0} \tx{i}{t} \tilde \alpha_i\agebs{i}{t} \geq \sum_{i \in \SrcSet : \ifbusy{i}{t}=0} \rateop{i} \tilde \alpha_i\agebs{i}{t}. 
\end{align}
Applying \eqref{max-weight-bound} to \eqref{drift-derive}, we obtain
\begin{align*}
     \Delta(\networkstate{t}) \leq & -\frac{1}{N}\sum_{i \in \SrcSet : \ifbusy{i}{t}=1} \tx{i}{t} \tilde \alpha_i\agebs{i}{t}+\frac{1}{N}\sum_{i\in \SrcSet} \tilde \alpha_i\nonumber\\
     &  -\frac{1}{N}\sum_{i\in \SrcSet: \ifbusy{i}{t}=0} \rateop{i} \tilde \alpha_i\agebs{i}{t}.
\end{align*} 
Since node $i$ with $\ifbusy{i}{t}=1$ is not selected for transmission, leading to $\tx{i}{t}=0$, the first term in the RHS above is $0$ and we have
\begin{align*}
     \Delta(\networkstate{t})&\leq \frac{1}{N}\sum_{i\in \SrcSet} \tilde \alpha_i -\frac{1}{N}\sum_{i; \ifbusy{i}{t}=0} \rateop{i} \tilde \alpha_i\agebs{i}{t}.
\end{align*} 
Rewriting the last sum above to \\
$$\frac{1}{N}\sum_{i\in \SrcSet} \rateop{i} \tilde \alpha_i\agebs{i}{t}\prtr{1-\ifbusy{i}{t}}$$ gives
\begin{align*}
     \Delta(\networkstate{t}) &\leq \frac{1}{N}\sum_{i\in \SrcSet} \tilde \alpha_i-\frac{1}{N}\sum_{i\in \SrcSet} \rateop{i} \tilde \alpha_i\agebs{i}{t} \\
      &+ \frac{1}{N}\sum_{i \in \SrcSet} \rateop{i} \tilde \alpha_i\agebs{i}{t}\ifbusy{i}{t}.
\end{align*} 
Summing the above over $t\in \prtc{1, 2, \dotsc, T}$ and dividing by $T$ yields

\begin{multline*}
     \frac{ \lfunction{\networkstate{T+1}}}{T}- \frac{ \lfunction{\networkstate{1}}}{T} \\
     \leq
      \frac{1}{N}\sum_{i\in \SrcSet} \tilde \alpha_i -\frac{1}{NT} \sum_{t=1}^{T} \sum_{i\in \SrcSet} \rateop{i} \tilde \alpha_i\agebs{i}{t} \\
     + \frac{1}{NT} \sum_{t=1}^{T} \sum_{i\in \SrcSet} \rateop{i} \tilde \alpha_i\agebs{i}{t}\ifbusy{i}{t}.
\end{multline*} 
Knowing $\frac{ \lfunction{\networkstate{t+1}}}{T}$ is positive, we manipulate the above to
\begin{multline*}
     \frac{1}{NT} \sum_{t=1}^{T} \sum_{i\in \SrcSet} \rateop{i} \tilde \alpha_i\agebs{i}{t} 
     \leq 
      \frac{ \lfunction{\networkstate{1}}}{T} + \frac{1}{N}\sum_{i\in \SrcSet} \tilde \alpha_i \\
     + \frac{1}{NT} \sum_{t=1}^{T} \sum_{i\in \SrcSet} \rateop{i} \tilde \alpha_i\agebs{i}{t}\ifbusy{i}{t}.
\end{multline*} 
Setting $\tilde \alpha_i =w_i/\rateop{i}$ in the above and applying the limit $T\to \infty$ with knowing that $ \lfunction{\networkstate{1}}$ is a constant, we get
\begin{multline}\label{max-weight-bound5}
     \lim_{T\to \infty}\frac{1}{NT} \sum_{t=1}^{T} \sum_{i\in \SrcSet} w_i\agebs{i}{t} \leq \frac{1}{N}\sum_{i\in \SrcSet} \frac{w_i}{\rateop{i}} \\
      + \lim_{T\to \infty} \frac{1}{NT} \sum_{t=1}^{T} \sum_{i\in \SrcSet} w_i\agebs{i}{t}\ifbusy{i}{t}.
\end{multline} 
Next, we bound the last term in \eqref{max-weight-bound5}. Fix node $i$. Consider packet transmissions during $T$ slots as in \eqref{timeslot-interval}. For $m>1$, the slot that follows the $(m-1)$th packet delivery from node $i$ has $\agebs{i}{t}=1$ and $\ifbusy{i}{t}=1$, and $\agebs{i}{t}\ifbusy{i}{t}$ evolves as $\prtc{1,2, \dotsc,\proctime{i}-1 ,0, \dotsc, 0}$ until the slot that the $m$th packet is transmitted. Therefore, the sum of $\agebs{i}{t}\ifbusy{i}{t}$ during the inter-deliver times between the $(m-1)$th and $m$th transmissions is $\proctime{i}(\proctime{i}-1)/2$, and the sum during the remaining $\remainslot{i}$ slots is at most $\proctime{i}(\proctime{i}-1)/2$. Since early slots before the first transmission, $\ifbusy{i}{t}$ is set to $0$, and we have

\begin{align*}
    & \frac{1}{T} \sum_{t=1}^{T} \agebs{i}{t} \ifbusy{i}{t}\\
    & \leq \frac{1}{T} \prts{ \prtr{\numpacket{i}{T}-1}\frac{\proctime{i}\prtr{\proctime{i}-1}}{2}+\frac{\proctime{i}\prtr{\proctime{i}-1}}{2}}\nonumber\\
    &=\frac{\proctime{i}\prtr{\proctime{i}-1}}{2}\frac{\numpacket{i}{T}}{T}.
\end{align*}
Taking limit $T\to \infty$ to the above and using the definition of the long-term transmission rate, we have
\begin{align*}
    \lim_{T\to \infty}\frac{1}{T} \sum_{t=1}^{T} \agebs{i}{t} \ifbusy{i}{t} \leq
     \frac{\proctime{i}\prtr{\proctime{i}-1}}{2} \ratepolicy{i}{\pimx}.
\end{align*}
Applying Lemma \ref{upper-bound-rate} to the above as $\pimx$ is drop-free yields
\begin{align*}
    \lim_{T\to \infty}\frac{1}{T} \sum_{t=1}^{T} \agebs{i}{t} \ifbusy{i}{t} &\leq\frac{\proctime{i}-1}{2}.
\end{align*}
Applying the above inequality to \eqref{max-weight-bound5}, we achieve
\begin{align*}
     &\lim_{T\to \infty}\frac{1}{NT} \sum_{t=1}^{T} \sum_{i\in \SrcSet}  w_i\agebs{i}{t}\\
     &\leq \frac{1}{N}\sum_{i\in \SrcSet} \frac{w_i}{\rateop{i}} + \frac{1}{N} \sum_{i\in \SrcSet}  w_i\frac{\proctime{i}-1}{2}.
\end{align*} 
Combining Lemma \ref{awsaoi-and-awsaoib} with $\pimx$ and the above inequality gives
\begin{align}\label{aoi-mx}
\AWSAoI{\pimx} \leq \frac{1}{N} \prts{ \sum_{i\in \SrcSet} \frac{w_i}{\rateop{i}}+ \sum_{i\in \SrcSet} w_i\frac{\proctime{i}-1}{2}+ \sum_{i\in \SrcSet} w_i\proctime{i} }
\end{align} 
Comparing \eqref{aoi-mx} with \eqref{bound}, the Max-Weight policy is $2$-optimal as 
\begin{align*}\label{aoi-mx2}
\AWSAoI{\pimx} \leq 2\prts{ \frac{1}{N} \sum_{i \in \SrcSet}\prtr{ \frac{ w_i}{2 \rateop{i}} +\frac{ w_i}{2} +w_i\proctime{i}}} 
\leq 2 \AWSAoI{*}.
\end{align*} 
Furthermore, if $\sum_{i \in \SrcSet} \frac{1}{\proctime{i}} \leq 1,$ then $\rateop{i} =\frac{1}{\proctime{i}}$. Hence, substituting $\rateop{i}=\frac{1}{\proctime{i}}$ to \eqref{aoi-mx} and comparing with \eqref{bound-special}, we have 
\begin{align*}
\AWSAoI{\pimx}&\leq\frac{1}{N}\prts{ \sum_{i\in \SrcSet} w_i\proctime{i} +\sum_{i\in \SrcSet}  w_i\frac{\proctime{i}-1}{2}+ \sum_{i\in \SrcSet} w_i\proctime{i} }\nonumber\\
&\leq \frac{5}{3}\prts{ \frac{1}{N} \sum_{i \in \SrcSet} \prtr{\frac{3}{2}w_i\proctime{i}+ \frac{w_i}{2}}} \leq \frac{5}{3} \AWSAoI{*}. 
\end{align*}
Thus, the Max-Weight policy is $5/3$-optimal if $\sum_{i \in \SrcSet} \frac{1}{\proctime{i}} \leq 1$.
\end{proof}

%% file: 6-SimulationResult.tex

\section{Simulation Results}

\label{sec:simulation}





This section evaluates the performance of \algonamepowertwo, \algonamebackoff, and Max-Weight policies in two aspects: AWSAoIP and maximum AoIP. We use the \textsf{CVXPY} package \cite{cvxpy} to solve the convex optimization in \eqref{bound-problem} for optimal transmission rates. For each scheduling policy, we simulate $10^6$ slots to ensure AWSAoIP converges.


\subsection{Average age of information processed}
\begin{figure}
    \centering
    \vspace{-1em}
    \includegraphics[width=\linewidth]{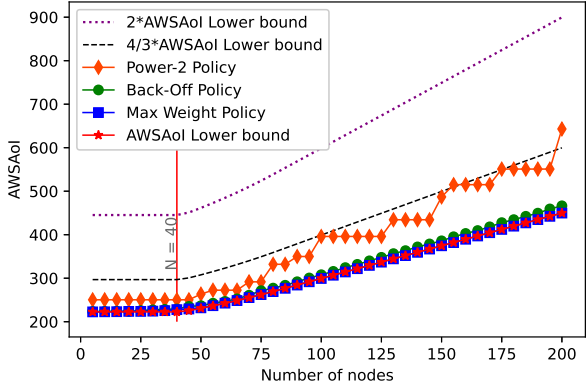}\vspace{-0.7em}
    \caption{\AWSAoIabbre s with varying numbers of nodes}
    \label{fig:varing_num_node}
\end{figure}

We evaluate \AWSAoIabbre~ under the three policies in networks with a varying number of nodes. We begin with a group of $5$ nodes with processing times $(24, 152, 70, 37, 54)$ and weights $(4.1, 7.2, 1.1, 3.0, 1.4)$. These processing times yield $\sum_{i=1}^5 \frac{1}{\proctime{i}}\approx 0.12$. In Figure \ref{fig:varing_num_node}, we simulate 40 network setups. Each setup contains a multiple of groups, and the largest setup has $5 \times 40 = 200$ nodes. 

The simulation result shows that both centralized policies--- \algonamebackoff and Max-Weight policies--- are near-optimal as their \AWSAoIabbre~ are close to the lower bound in \eqref{bound-problem}. For large networks, Max-Weight policy slightly outperforms the \algonamebackoff policy. The \algonamepowertwo policy trades performance for the simple distributed algorithm, yet it still achieves $2$-optimal. The \AWSAoIabbre s from these policies are flat until $N = 40$ because $\sum_{i=1}^N \frac{1}{\proctime{i}} \leq 1$, so the systems are processor bound. After $N = 40$, the systems are communication bound and \AWSAoIabbre~ increases linearly with the number of nodes.


\subsection{Maximum age of information processed}
\begin{figure}
    \centering
    \vspace{-1em}
    \includegraphics[width=\linewidth]{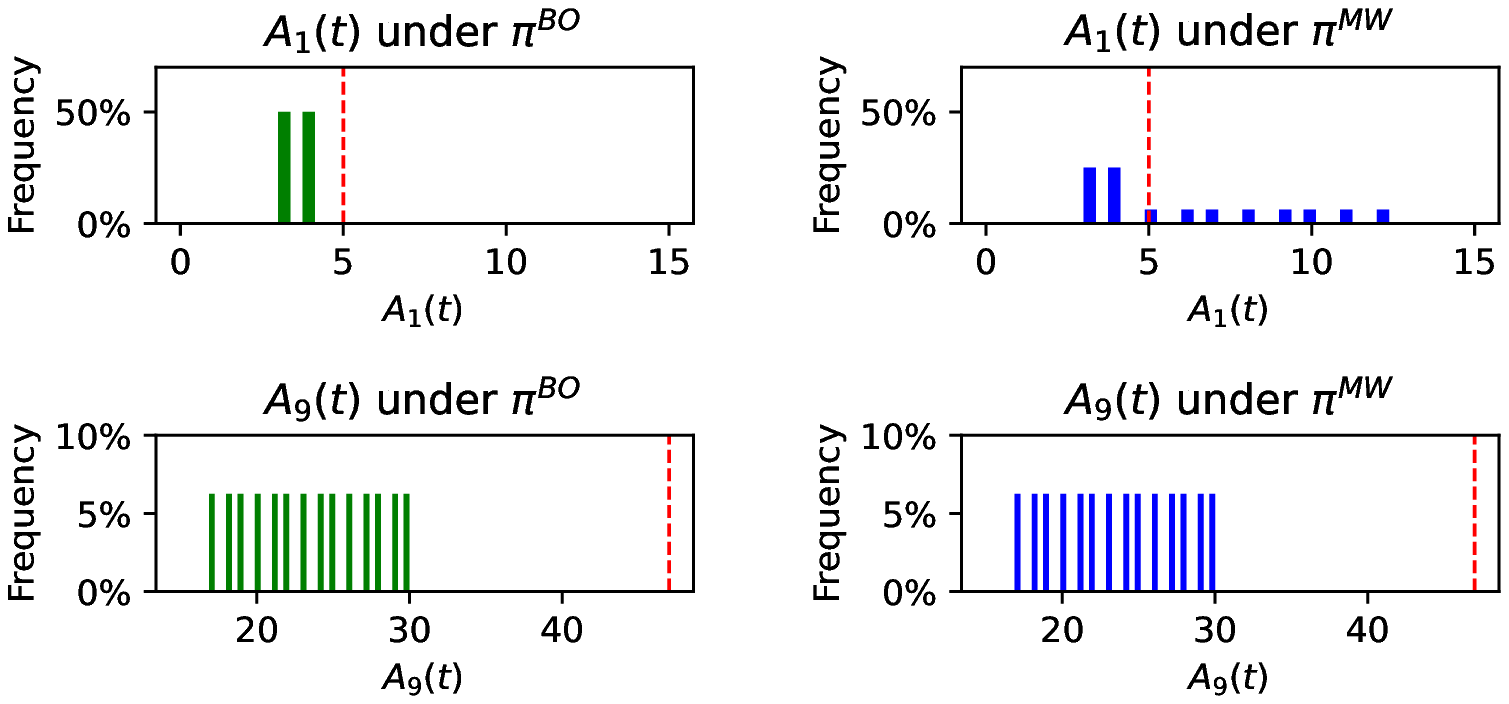}\vspace{-1.5em}
    \caption{ The distributions of $\procage{1}{t}$ with $\proctime{1}=2$ and $\procage{9}{t}$ with $\proctime{1}=16$ under the \algonamebackoff and Max-Weight policies. The red dashed lines represent the bounds on the maximum AoI derived in Corollary \eqref{max-age}.}
    \label{fig:distribution}
\end{figure}

We illustrate that the \algonamebackoff policy guarantees a bound on the maximum AoIP as in Corollary \ref{max-age}. Consider a network setup consisting of $9$ nodes, where $\proctime{1}=2$ and $\proctime{i}=16$ for $i \in \prtc{2, 3, \dotsc, 9}$ and all weights are $1$. The $\AWSAoI{\pibackoff}$ and $\AWSAoI{\pimx}$ in this setup are $199.5$ and $202.0$ respectively. Figure \ref{fig:distribution} shows the frequencies of $\procage{1}{t}$ and $\procage{9}{t}$ under both policies, and the red dash lines are the maximum AoIP from Corollary \ref{max-age}. The \algonamebackoff policy keeps all AoIPs within the maximum. Furthermore, it is possible to construct a setup for the Max-Weight policy that the frequency of event $\procage{1}{t}=y$ is non-zero while $\rateop{1}=0.5$ for any positive integer $y$, and the maximum AoIP is unbounded linearly by $c(1/\rateop{i})$ for any constant $c$.

%% file: 7-Conclusion.tex
\section{Conclusions and Future Research}
\label{sec:conclusion}

This paper considered a wireless network with nodes transmitting unprocessed packets to their corresponding non-preemptive processors at the base station over a shared wireless channel. We designed three scheduling policies: \algonamepowertwo, \algonamebackoff, and \algonamemx to minimize the \textAWSprocessaoi~ of the system. All policies are $2$-optimal and have different advantages. The \algonamepowertwo policy is simple and distributed. The \algonamebackoff policy is near-optimal and has a maximum AoIP bound guarantee. The \algonamemx policy achieves closest to the system's lower bound. We confirm the results through simulations.


Our subsequent work will consider a system model with unreliable wireless channels, where we have some preliminary results. Future research can consider the periodic generation of packets, minimum throughput constraints, and random processing times.


%% file: 8-Appendix.tex

\appendix
\label{sec:Appendix}

\section{Proof of Lemma 6} \label{no-zero-appendix}
The proof of Lemma 6 requires defining new definitions and proving the other three lemmas. Therefore, we begin with the requirements and prove Lemma 6 at the end of this appendix.





First, we introduce several definitions.  Define 
\begin{align*}
    \whichnodesent{t}=\begin{cases} 
      i & , \tx{i}{t}=1 \text{ for some } i \in \SrcSet;\\
      0 & , \tx{i}{t}=0 \text{ for all } i \in \SrcSet
   \end{cases}
\end{align*}
to indicate which node is selected to transmit a packet during slot $t$ as shown in Table \ref{tab:example_case_appendix}. That is,  if $\whichnodesent{t}\neq 0,$ then $\countdown{\whichnodesent{t}}{t}$ is the countdown time of the node selected to transmit a packet in slot $t.$

We define the \emph{wait gap} $\waitgapat{t}$ as the set of the time slots that the node selected to transmit in slot $t$ has waited since it stopped being backed off. For example, in Table \ref{tab:example_case_appendix}, node $i=3$ is scheduled for a transmission in slot $t=12$, and we have that $\waitgapat{12}=\prtc{12, 11}$. Generally, it holds that 
\begin{align*} \waitgapat{t}=\begin{cases} 
       \big\{ t, t-1, \dotsc,t-\proctimeBO{\whichnodesent{t}} &+\countdown{\whichnodesent{t}}{t} \big\}\\
      &,\sum_{n \in \SrcSet} \tx{n}{t} =1;\\
      \emptyset &,\sum_{n \in \SrcSet} \tx{n}{t} =0.
   \end{cases}
\end{align*}



\begin{table}[]
\resizebox{\columnwidth}{!}{%
\begin{tabular}{|c|c|c|c|c|c|c|c|c|c|c|c|c|}
\hline
t       & 1 & 2 & 3 & 4 & 5 & 6 & 7 & 8 & 9 & 10 & 11 & 12 \\ \hline
$S(t)$  &  1 & 2  &  1  &  3 &  1 &  2 &  1 &  0 &  1  &  2  &  1  &  3     \\ \hline
$C_1(t)$ &$\cancel{2}$&$\infty$&$\cancel{2}$&$\infty$&$\cancel{2}$&$\infty$&  $\cancel{2}$ & $\infty$  &   $\cancel{2}$ &  $\infty$  &  $\cancel{2}$  &   $\infty$     \\ 
$C_2(t)$ &  3 &$\cancel{2}$&$\infty$&$\infty$&3&  $\cancel{2}$ & $\infty$  & $\infty$  &  3 &  $\cancel{2}$  & $\infty$   &  $\infty$    \\ 
$C_3(t)$ &  7 & 6 & 5  &$\cancel{4}$&$\infty$&$\infty$&$\infty$&$\infty$&$\infty$& $\infty$   &  7  &  $\cancel{6}$     \\ \hline
$\ifbusybackoff{3}{t}$  &  0 & 0  &  0  &  0 &  1 &  1 &  1 & 1 &  1 &  1  &  0  &  0     \\ \hline
$\ifbusy{3}{t}$  &  0 & 0  &  0  &  0 &  1 &  1 &  1 &  0 &  0  &  0  &  0  &  0     \\ \hline
\end{tabular}
}
\\
   \caption{Extension of Table \ref{tab:example_case} with $S(t)$. 
Also, we have $\waitgapat{4}=\prtc{1,2,3,4},\waitgapat{3}=\{3\} $, $\waitgapat{2}=\prtc{1,2},$ $\waitgapat{1}=\prtc{1},$ and $\waitgapat{8}=\emptyset.$  Furthermore, $(\waitgapat{4},\waitgapat{2},\waitgapat{1})$ is a \connectedchain, but $(\waitgapat{4},\waitgapat{3},\waitgapat{2},\waitgapat{1})$ is not since $\waitgapat{3}\cap \waitgapat{2} = \emptyset$.}
 
\label{tab:example_case_appendix}
\end{table}



We then define a \emph{\connectedchain} as an order of wait gaps $\waitgapat{t_1},\waitgapat{t_2}, \dotsc, \waitgapat{t_h}$ where $t_1 > t_2 > \dotsc > t_h$ and $\waitgapat{t_i}\cap \waitgapat{t_{i+1}} \neq \emptyset$ for $i=1, \dotsc, h-1$. The last condition and the definition of wait gap implies that a connected chain is a duration that has a packet transmission in every slot. Finally, define a \emph{\firstzeroconnectedchain} as an order of $\waitgapat{t_1},\waitgapat{t_2}, \dotsc, \waitgapat{t_h}$ that is a \connectedchain and $\countdown{\whichnodesent{t_1}}{t_1}=0$, and there is no slot $t$ for $0<t<t_1$ that $\countdown{\whichnodesent{t}}{t}=0.$



We first assume that a \firstzeroconnectedchain~ exists, which later leads to a contradiction and there is no such a chain. Suppose a \firstzeroconnectedchain~ $\firstzeroconnectedchainexample = \prtr{\waitgapat{t_1},\dotsc, \waitgapat{t_k}}$ exists such that $\countdown{i^*}{t_1}=0 $ for some node $i^*$. We denote $\chainsetfzcg =\waitgapat{t_1}\cup  \dotsc \cup \waitgapat{t_k}$ and $\lengthchain=\abs{\chainsetfzcg}$. We assume that $\firstzeroconnectedchainexample$ is with the longest length. This chain leads to the following three lemmas, as follow:



\begin{lemma}[countdown time in the chain]
\label{at-most-l}
For the \firstzeroconnectedchain~ $\firstzeroconnectedchainexample$, it holds that $\countdown{\whichnodesent{t}}{t}\leq t_1-t$ for every  $t \in \chainsetfzcg$.  
\end{lemma}
\begin{proof}
Let $P(j)$ be the statement that $\countdown{\whichnodesent{t}}{t}\leq t_1-t$ for all $t \in \waitgapat{t_j}$, where $j=1,2, \dotsc,k.$

\textbf{Basic Step:}  Since $\countdown{\whichnodesent{t_1}}{t_1}=0$, we then have that $\countdown{\whichnodesent{t_1}}{t}=t_1-t$ for every $t \in \waitgapat{t_1}$ by the evolution of the countdown time in in (\ref{countdown-def}). Since node $\whichnodesent{t_1}$ does not get to transmit a packet during slot $t \in  \waitgapat{t_1}/\prtc{t_1},$ the countdown time of the selected node in slot $t$ must not exceed the countdown time of node $\whichnodesent{t_1}.$ That is, 
\begin{align}
    \countdown{\whichnodesent{t}}{t} \leq \countdown{\whichnodesent{t_1}}{t}=t_1-t
\end{align}
for all $t \in  \waitgapat{t_1}/\prtc{t_1}.$ Combining with the fact that $\countdown{\whichnodesent{t_1}}{t_1}=0,$ $P(1)$ is true.\\\\
\textbf{Inductive Step:} Let $q$ be a natural number such that $q<k$ and $P(q)$ is true. That is, for every $t\in \waitgapat{t_q},$   $\countdown{\whichnodesent{t}}{t}\leq t_1-t.$

Since $\waitgapat{t_q}\cap \waitgapat{t_{q+1}} \neq \emptyset,$ we can see that $t_{q+1} \in \waitgapat{t_q}$ and 
\begin{align}\label{inductive-1}
\countdown{\whichnodesent{t_{q+1}}}{t_{q+1}}\leq t_1-t_{q+1}.
\end{align}
 Using the similar approach as in the Basic Step, the evolution of $\countdown{\whichnodesent{t_{q+1}}}{t}$ in (\ref{countdown-def}) implies that 
\begin{align}\label{inductive-2}
\countdown{\whichnodesent{t_{q+1}}}{t}= \countdown{\whichnodesent{t_{q+1}}}{t_{q+1}}+\prtr{t_{q+1}-t}
\end{align}
for all $t \in  \waitgapat{t_{q+1}}.$

Since  $\whichnodesent{t_{q+1}}$ does not get to transmit a packet during slot $t \in  \waitgapat{t_{q+1}}/\prtc{t_{q+1}},$ the countdown time of the selected node must not exceed the countdown time of node $\whichnodesent{t_{q+1}}.$ That is, combining with (\ref{inductive-1}) and (\ref{inductive-2}), we have
\begin{align}
    \countdown{\whichnodesent{t}}{t} &\leq \countdown{\whichnodesent{t_{q+1}}}{t} \nonumber\\
    &\leq  \prtr{t_1-t_{q+1}}+\prtr{t_{q+1}-t}=t_1-t
\end{align}
for all $t \in  \waitgapat{t_{q+1}}/\prtc{t_{q+1}}.$ Thus, combining with (\ref{inductive-1}), $P(q+1)$ is true.

By the proof of induction, we can conclude that, for every $t \in \chainsetfzcg,$  $\countdown{\whichnodesent{t}}{t}\leq t_1-t.$
\end{proof}

\begin{lemma}[processing time shorter than the chain]
\label{smaller-than-length}
For the \firstzeroconnectedchain~ $\firstzeroconnectedchainexample$, it holds that $\proctimeBO{\whichnodesent{t}} \leq \lengthchain-1$ for every $t \in \chainsetfzcg$.
\end{lemma}
\begin{proof}
Assume that there exists some $t' \in \chainsetfzcg$ such that $\proctimeBO{\whichnodesent{t'}} \geq \lengthchain.$ There must be some $q\in\prtc{1,2, \dotsc,k}$ that $t' \in \waitgapat{t_q}$ and $t' \neq t_q$ 
We know that 
\begin{align}\label{smaller-1}
    \abs{\waitgapat{t'}}&=\abs{ \prtc{t',t'-1, \dotsc,t'-\proctimeBO{\whichnodesent{t'}} +\countdown{\whichnodesent{t'}}{t'}}} \nonumber\\
    &=\proctimeBO{\whichnodesent{t'}} -\countdown{\whichnodesent{t'}}{t'}+1.
\end{align}
Applying Lemma \ref{at-most-l} and the assumption that  $\proctimeBO{\whichnodesent{t'}} \geq \lengthchain$ to (\ref{smaller-1}), we have
\begin{align}
    \abs{\waitgapat{t'}}&\geq \lengthchain -\prtr{t_1-t'}+1.
\end{align}
Thus, we consider $G' =\waitgapat{t_1}\cup \waitgapat{t_2}\cup \dotsc \cup \waitgapat{t_q}\cup \waitgapat{t'}.$ Since $t' \in \waitgapat{t_q},$ we can easily prove that $\prtr{\waitgapat{t_1},\waitgapat{t_2}, \dotsc,\waitgapat{t_q} ,\waitgapat{t'}}$ is a \firstzeroconnectedchain. However, we know that $\prtc{t_1,t_1-1, \dotsc,t'+1} \subseteq G'/\waitgapat{t'}$ and 
\begin{align}
    \abs{G'}&= \abs{G'/\waitgapat{t'}}+\abs{\waitgapat{t'}}\nonumber\\
    &\geq \prtr{t_1-t'}+ \abs{\waitgapat{t'}}\nonumber\\
    &\geq \prtr{t_1-t'} +\lengthchain -\prtr{t_1-t'}+1=\lengthchain+1,
\end{align}
which is longer than $\lengthchain=\abs{\chainsetfzcg}.$ This causes a contradiction since $\firstzeroconnectedchainexample$ is the longest \firstzeroconnectedchain. Therefore, $\proctimeBO{\whichnodesent{t}} \leq \lengthchain-1$ for every $t \in \chainsetfzcg.$
\end{proof}

Lemmas \ref{at-most-l} and \ref{smaller-than-length} are used to prove the following lemma.

\begin{lemma}[maximum transmissions in the chain]
\label{number-floor}
For the \firstzeroconnectedchain~ $\firstzeroconnectedchainexample$, the number of transmissions from node $i$ during $t \in  \chainsetfzcg \backslash \prtc{t_1}$ is at most $\floor{\prtr{\lengthchain-1}/\proctimeBO{i}}$, for every $i\in \SrcSet.$

\end{lemma}

\begin{proof}

For node $i$ that $\proctimeBO{i}> \lengthchain-1,$ it does not get to transmit a packet within $t \in  \chainsetfzcg \backslash \prtc{t_1}$ at all by Lemma \ref{smaller-than-length}. In other words, it transmits at most $\floor{\prtr{\lengthchain-1}/\proctimeBO{i}}=0$. Now consider node $i$ such that $\proctimeBO{i}\leq \lengthchain-1.$

If $\lengthchain-1$ is divisible by $\proctimeBO{i},$ then we can partition these $\lengthchain-1$ slots into $\frac{\lengthchain-1}{\proctimeBO{i}}$ intervals of consecutive $\proctimeBO{i}$ slots. For node $i,$ the inter-delivery time is always at least $\proctimeBO{i}.$ Thus, each interval can transmit at most $1$ packet from node $i.$ That is, during these $\lengthchain-1$ slots, there are at most $\frac{\lengthchain-1}{\proctimeBO{i}}=\floor{\prtr{\lengthchain-1}/\proctimeBO{i}}$ transmissions from node $i.$

If $\lengthchain-1$ is not divisible by $\proctimeBO{i},$ we let $z_i$ be the remainder where $1\leq z_i<\proctimeBO{i}.$ Again, we partition $\chainsetfzcg \backslash \prtc{t_1}$ into $\floor{\prtr{\lengthchain-1}/\proctimeBO{i}}+1$ intervals where the last interval is  $\prtc{t_1,t_1+1, \dotsc,t_1+z_i-1}$ of length $z_i$ and the rest are $\floor{\prtr{\lengthchain-1}/\proctimeBO{i}}$ intervals where each interval is a consecutive $\proctimeBO{i}$ slots. Using the similar argument, we can transmit at most $\floor{\prtr{\lengthchain-1}/\proctimeBO{i}}+1$ where each interval has one transmission from node $i.$ If we get exact $\floor{\prtr{\lengthchain-1}/\proctimeBO{i}}+1$ transmissions, then node $i$ must transmit one packet during the second-to-last interval and another one during the last interval. However, when we transmit a packet during the second-to-last interval, $\countdown{i}{t}$ will be set to $\infty$ for $\proctimeBO{i}-1$ slots, then become $\proctimeBO{i},$ and keep decreasing by $1$ in every slot until node $i$ transmits another packet. We can easily check that, during the last interval, $\countdown{i}{t}$ will always greater than $t_1-t.$ However, by Lemma \ref{at-most-l}, node $i$ would not be able to transmit a packet during the last $z_i$ slots, contradicting the assumption that node $i$ transmits a packet during the last interval. Therefore, for every $i \in \SrcSet$, there are at most $\floor{\prtr{\lengthchain-1}/\proctimeBO{i}}$ transmissions from node $i$ within $t \in  \chainsetfzcg \backslash \prtc{t_1}$.
\end{proof}
Now, we will show that if such \firstzeroconnectedchain~ $\firstzeroconnectedchainexample$ exists, it leads to a contradiction. 
\begin{customthm}{6}[No-zero] 
If $\sum_{i=1}^N \frac{1}{\proctimeBO{i}}\leq 1$, then we have that $\countdown{i}{t}\geq 1$ for every $i\in \SrcSet$ and every $t \in \prtc{1,2, \dotsc}$.  \end{customthm}

\begin{proof}
We assume that  there exists some $t^*$ and $i^*$ that $\countdown{i^*}{t^*}=0 $, so we must be able to detect  a \firstzeroconnectedchain \ $ \firstzeroconnectedchainexample =\prtr{\waitgapat{t_1},\waitgapat{t_2}, \dotsc, \waitgapat{t_k}}$ where $t^* = t_1$. To be concise, we denote $\chainsetfzcg =\waitgapat{t_1}\cup \waitgapat{t_2}\cup \dotsc \cup \waitgapat{t_k}$ and $\lengthchain=\abs{\chainsetfzcg}.$ Since there might be multiple \firstzeroconnectedchain s, we assume that $\firstzeroconnectedchainexample$ is the \firstzeroconnectedchain~ with the largest $\lengthchain.$

By Lemma \ref{number-floor}, the total number of transmissions during interval $\chainsetfzcg \backslash \prtc{t_1}$ is at most
\begin{align} \label{no-zero:1}
    \sum_{i\in \SrcSet} \floor{\prtr{\lengthchain-1}/\proctimeBO{i}}&\leq \sum_{i\in \SrcSet} \prtr{\lengthchain-1}/\proctimeBO{i}\nonumber\\
    &=\sum_{i\in \SrcSet} \prtr{\lengthchain-1}/\ceiling{\frac{1}{\rateop{i}}}\nonumber\\
    &\leq \sum_{i\in \SrcSet} \prtr{\lengthchain-1}\rateop{i}\nonumber\\
    &\leq \lengthchain-1.
\end{align}

Since the chain is connected, there must be exactly  $\lengthchain -1$ transmissions  during during interval. Thus, the inequalities in (\ref{no-zero:1}) must be equalities. That is, $\lengthchain -1$ is divisible by $\proctimeBO{i}$ for every $i\in \SrcSet$, and node $i$ must exactly transmit $\floor{\prtr{\lengthchain-1}/\proctimeBO{i}}= \prtr{\lengthchain-1}/\proctimeBO{i}$ packets during this interval. We then consider node $i^*=\whichnodesent{t_1}.$ Since $\countdown{i^*}{t_1}=0,$ this means that node $i^*$ cannot transmit during interval $\prtc{t_1,t_1-1, \dotsc ,t_1-\proctimeBO{i^*}}.$ That is, during interval $\chainsetfzcg \backslash \prtc{t_1},$ node $i^*$ can transmit at most $\ceiling{\frac{\lengthchain-\proctimeBO{i^*}-1}{\proctimeBO{i^*}}}=\frac{\lengthchain-1}{\proctimeBO{i^*}}-1,$ resulting in a contradiction. Therefore, $\countdown{i}{t}\geq 1$ for every $i\in \SrcSet$ and $t=1,2, \dotsc$
\end{proof}





